\documentclass[12pt,draftclsnofoot,onecolumn]{IEEEtran}
\usepackage{amsmath}
\usepackage{amssymb}
\usepackage{algorithm}
\usepackage{algorithmicx}
\usepackage{algpseudocode}
\usepackage[T1]{fontenc}
\usepackage{color}
\usepackage{times}
\usepackage{cite}
\usepackage{caption}
\usepackage{graphicx}
\usepackage{float} 
\usepackage{subcaption}
\usepackage{setspace}
\usepackage{stfloats}

\begin{document}
	
	\makeatletter
	\renewcommand{\maketag@@@}[1]{\hbox{\m@th\normalsize\normalfont#1}}%
	\makeatother
	
	\newenvironment{sequation}{\begin{equation}\small}{\end{equation}}
	\newcounter{TempEqCnt}
	\setlength{\abovedisplayskip}{5pt}
	\setlength{\belowdisplayskip}{3pt}
	
\title{To Reflect or Not To Reflect: On-Off Control and Number Configuration for Reflecting Elements in RIS-Aided Wireless Systems}
\author{Hao Xie,  Dong Li,~\IEEEmembership{Senior Member,~IEEE} 
	\thanks{H. Xie and D. Li are with the School of Computer Science and Engineering, Macau University of Science and Technology, Avenida Wai Long, Taipa, Macau 999078, China (e-mails: 3220005631@student.must.edu.mo,  dli@must.edu.mo).}
\vspace{-30pt}} 

\maketitle

\begin{abstract}
Reconfigurable intelligent surface (RIS) has been regarded as a promising technique due to its high array gain, low cost, and low power. However,  the traditional passive RIS suffers from the ``double fading'' effect, which has become a major bottleneck in restricting the performance of passive RIS-aided communications. Fortunately, active RIS can alleviate this problem since it can adjust the phase shift and amplify the received signal simultaneously. Nevertheless, a high beamforming gain often requires a large number of reflecting elements, which leads to non-negligible power consumption, especially for the active RIS. Thus, one challenge is how to improve the scalability of the RIS and the energy efficiency. Different from the existing works where all reflecting elements are activated, we propose a novel element on-off mechanism where reflecting elements can be flexibly activated and deactivated. To achieve a tradeoff between the transmission rate and energy consumption, two different optimization problems for passive RIS and active RIS are formulated by maximizing the total energy efficiency, where the constraints of the maximum power of users and the RIS, the minimum transmission rate, the element on-off factor, and the unit moduli of passive elements are taken into account. In light of the intractability of the formulated problems, we develop two different alternating optimization-based iterative algorithms by combining quadratic transform, variable substitution, and the successive convex approximation method to obtain sub-optimal solutions. Furthermore, in order to gain more insight into problems, we consider special cases involving transmission rate maximization problems for given the same total power budget, and respectively analyze the number configuration for passive RIS and active RIS. Simulation results verify that the proposed algorithms outperform existing algorithms, and reflecting elements under the proposed algorithms can be flexibly activated and deactivated.
\end{abstract}


\section{Introduction}
\subsection{Backgroud}
With the increasing prevalence and rapid development of Internet of Things (IoT) technologies, all kinds of passive tags, industrial sensors and controllers, and wireless devices are continuously  integrated into IoT ecosystems, significantly enhancing the capabilities of  automatic identification, signal processing, information transmission and  exchange,  and real-time tracking\cite{a1,a2}. However, traditional wireless communication technologies mainly rely on components such as high-power antennas and amplifiers. With the increasing numbers of IoT devices and the growing demand for high throughput, there will be inevitably significant energy consumption. Therefore, effective technologies are urgently needed to achieve low energy consumption, high efficiency, and sustainable development of wireless communication networks.

Reconfigurable intelligent surface (RIS) has recently emerged as a promising technology for enhancing the capacity and coverage of wireless communication systems, which aligns with the concept of green and sustainable wireless communication networks since the RIS is essentially a low-power device\cite{b1,b2}. Specifically, the RIS is composed of a number of passive reflecting elements that can adjust the phase and amplitude of the incident electromagnetic waves. By employing passive reflecting elements, RISs can alter the propagation characteristics of wireless signals and achieve beamforming, spatial focusing, and interference mitigation, thereby expanding the signal coverage area for signal blind zone and enhancing the signal strength for hotspots. However, the signal undergoes cascaded channels, i.e.,  from the source to the RIS and from the RIS to the destination, which will result in the ``double fading'' effect. This effect significantly deteriorates the quality of signal transmission and limits the performance of the RIS in practical applications.

There have been a number of research efforts that have been paid to overcome the above challenges. First, one solution to overcome the ``double fading'' effect is to deploy a large number of reflecting elements, which makes the beam more focused and strengthened, thereby achieving a higher passive beamforming gain \cite{c1,c2}. This gain can partially compensate for the ``double fading'' effect and improve the quality and coverage of signal transmission. However, in practical systems, the number of passive elements is limited due to the high channel acquisition overhead\cite{c3}. This is because the required pilot overhead of existing cascaded channel estimation schemes is proportional to the number of passive elements. Therefore, increasing the number of passive elements results in a higher overhead for acquiring the channel state information. Second,  another way to overcome this effect is to deploy the RIS near the transmitter and receiver, which can reduce the transmission distance of the signal and reduce the loss during transmission\cite{c4}. Besides, there are currently some works on optimizing the deployment of the RIS\cite{c5,c5-1}. However, it is still difficult to obtain the optimal deployment in a more complicated environment with different device locations. Third, deploying multiple passive RISs can also alleviate the ``double fading'' effect\cite{c6}. Multiple RISs can reflect and adjust signals at different angles and distances, compensating for the ``double fading'' effect and providing a more stable signal to the receiver\cite{c7}. However, the cooperative control of multiple RISs will increase the complexity and adjustment difficulty of the system\cite{c8,c9}. The multiple reflections of interference will also impose a negative impact on the system performance. Thus, in light of the above discussion, more efficient technologies are still imperative to overcome this challenge.

Active RIS has recently emerged to enhance network performance, improve energy efficiency, and enable new communication services\cite{d1}. Different from the passive RIS, the active RIS integrates active components, such as reflection-type amplifiers and phase shifters, to dynamically adjust the signal reflection and transmission properties. Thus, the active RIS can adjust the phase shift and amplify the received signal simultaneously, and can eliminate signal attenuation and the ``double fading'' effect. So far, the active RIS has been applied  in several scenarios, including non-orthogonal multiple access\cite{d2}, wireless-powered communication network\cite{d3}, physical layer security\cite{d4}, satellite\cite{d5}, and simultaneous wireless information and power transfer\cite{d6}. These works have shown promising results in achieving various performance indicators, such as maximizing the sum rate, minimizing energy consumption, and enhancing the secrecy rate. Furthermore, several practical issues have also been considered in the design and implementation of the active RIS, such as power consumption and hardware constraints\cite{d7}.

\subsection{Motivation and Contributions}
Although extensive studies have been conducted on the passive RIS and the active RIS in recent years, some fundamental issues still remain unsolved.  One of the major misconceptions in the literature is the assumption that the power consumption of the RIS can be neglected due to its low power consumption. The power consumption of a single reflecting element is indeed low compared to relay and massive multiple-input-multiple-output. However, the sum of power consumption may be high when a large number of reflecting elements are applied\cite{d8}. Especially in the active RIS, the active RIS typically has higher cost and power consumption compared to the passive RIS\cite{d9}. However, the spectrum efficiency requirement is generally considered in existing works. In this case, it is more worthwhile to evaluate the energy efficiency performance of the RIS rather than the spectral efficiency. Moreover, it is worth noting that the energy efficiency is quite low when a large number of reflecting elements are employed\cite{d10}. Thus, a natural question arises: How to improve the scalability of the RIS and the energy efficiency while ensuring the quality of service (QoS) in passive/active RIS-aided wireless communication networks?


%
%
%

Motivated by the above observations, we propose a novel element on-off mechanism to improve the scalability of the RIS. The basic idea behind this mechanism is that, by selectively activating and deactivating the reflecting elements, the energy consumption of the RIS can be optimized/reduced. For instance, when a low power budget is required, activating only a portion of the reflecting elements can achieve the desired performance, which helps to reduce operational costs. Note that the element on-off mechanism has received little attention, and \cite{d8,e1} are highly related to this work regarding the element on-off mechanism. However,  \cite{d8,e1} investigate the element on-off mechanism under a given number of reflecting elements working on the ``on'' mode, which is significantly different from the element on-off mechanism design in this work (without any prior restriction on either ``on'' or ``off'' mode) regarding the problem formulation and resulting solution methodology. Besides, followed by the considered on-off control mechanism, the resulting number configuration for the active reflecting elements and the extension to the active RIS are worth investigating, which are not presently available in existing works. In this paper, we investigate and analyze the element on-off mechanism and evaluate the impact of the mechanism on  the energy efficiency performance in the passive RIS and active RIS-aided communication networks. The contributions of this work are summarized as follows:
\begin{itemize}
	\item We consider uplink transmission for RIS-aided communication networks.  To achieve a tradeoff between the sum rate and power consumption, we analyze  the impact of the element on-off mechanism on the energy efficiency performance for the passive RIS and the active RIS, respectively. Specifically, the total energy efficiency is maximized by jointly optimizing the transmit power, the receive beamforming vector, the phase-shift matrix, the element on-off factor, and the amplification factor subject to the maximum power constraint of users and the RIS, the minimum transmission rate constraint, the element on-off factor constraint, and the unit-modulus constraints of the passive RIS.
	\item In light of the intractability of the formulated problems, we develop two alternating optimization (AO)-based algorithms that adopt the variable substitution method, the quadratic transform method, and the successive convex approximation (SCA) method to obtain the corresponding sub-optimal solutions. In particular, the closed-form solutions of the receive beamforming vectors and transmit power are obtained based on the linear minimum-mean-square-error (MMSE) detection and the Karush-Kuhn-Tucker (KKT) conditions, respectively.
	\item To gain more insight into the number configuration of reflecting elements, we reformulate the energy efficiency optimization problems for the passive RIS and the active RIS  to transmission rate maximization problems for a given same total power budget, respectively. Specifically, we derive the closed-form solutions of number configuration for the active RIS and the passive RIS, and compare the performance of the element on-off mechanism in the two types of RIS. It is noted that the proposed on-off control mechanism and the number configuration can be designed in an offline manner.
	\item Simulation results verify the flexibility of the proposed algorithms, which can flexibly activate or deactivate reflecting elements based on external conditions, and thus outperforms the baseline algorithms in terms of energy efficiency.
\end{itemize}

\textit{Notations:} $\mathbb{E}(\cdot)$ and $\mathbb{C}^{N \times M}$ are the statistical expectation and $N \times M$ dimensional complex-valued matrix, respectively. ${\rm diag}(\cdot)$, ${\rm Re}(\cdot)$, and ${\rm Tr}(\cdot)$ denote the diagonalization, real part, and the trace, respectively. $x$, $\boldsymbol{\rm x}$, and $\boldsymbol{\rm X}$ denote the scalar, the vector, and the matrix, respectively. $|\cdot|$, $\|\cdot\|$, and $\|\cdot\|_F$ denote the absolute value,  the Euclidean norm, and the Frobenius norm, respectively. $\boldsymbol{{\rm I}}_M$ and $\boldsymbol{0}$ denote the $M \times M$ identity matrix and all-zero matrix, respectively. $\mathcal{CN}(\mu,\sigma^2)$ denotes the distribution of a circularly symmetric complex Gaussian random variable with mean $\mu$ and variance $\sigma^2$. $\boldsymbol{\rm X}^T$, $\boldsymbol{\rm X}^H$, $\boldsymbol{\rm X}^*$, and $[\boldsymbol{{\rm X}}]_{i,j}$ are the transpose, the conjugate transpose, conjugate, and the $(i,j)$-th element of matrix $\boldsymbol{\rm X}$, respectively. $\boldsymbol{\rm X}\succeq \boldsymbol{0}$ indicates that $\boldsymbol{\rm X}$ is a positive semidefinite matrix. $\left \lfloor \cdot \right \rfloor$ and $\left \lceil \cdot \right \rceil$ denote the floor and the ceil operations, respectively. $\arg(\cdot)$ denotes the angle of a complex number.

\subsection{Organization}
The remainder of this paper is structured as follows: The system model is presented in Section  \uppercase\expandafter{\romannumeral2}. The element on-off algorithm for the passive RIS is presented in Section \uppercase\expandafter{\romannumeral3}. Section \uppercase\expandafter{\romannumeral4} introduces the element on-off algorithm for the active RIS. Section \uppercase\expandafter{\romannumeral5} gives the number configuration of reflecting element.  Section \uppercase\expandafter{\romannumeral6} gives simulation results. The paper is concluded in Section \uppercase\expandafter{\romannumeral7}.

\section{System Model}
\begin{figure*}\vspace{-30pt}
	\centering
	\includegraphics[width=5.5in]{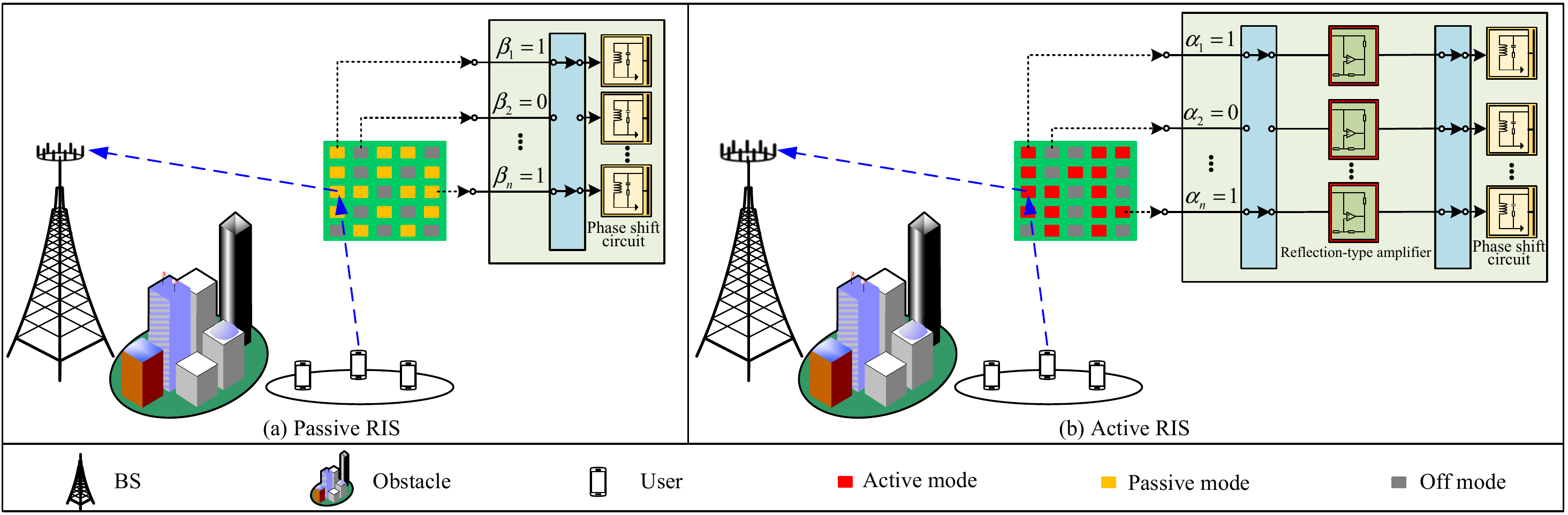}
	\caption{A RIS-aided wireless communication with element on-off.}
	\label{fig0}
\end{figure*}

As illustrated in Fig. \ref{fig0}, we consider a RIS-aided uplink communication network consisting of a base station (BS) with $M$ antennas, a RIS, and $K$ single-antenna users, where the direct link between the BS and users is blocked and can be ignored due to the unfavorable propagation conditions\cite{h1}. The RIS equipped with $N$ reflecting elements is placed in the cell to assist the uplink information transmission of users. To ensure the high energy efficiency of the system, reflecting elements can switch modes between on (active) mode and off mode for the active RIS, and on (passive) mode and off mode for the passive RIS. 
\subsection{Signal Transmission Model for Passive RIS}
The received signal at the BS is given by
\begin{sequation}\label{eq1}
\begin{split}
\boldsymbol{\rm y}^{\rm pas}
=\sum\limits_{k=1}^K\boldsymbol{\rm H}^H\boldsymbol{\rm B}\boldsymbol{\rm \Theta}\boldsymbol{\rm h}_{{\rm r}, k}\sqrt{p_k}s_k  + \boldsymbol{\rm n},
\end{split}
\end{sequation}
where $p_k$ and $s_k$ denote the transmit power and the information signal of the $k$-th user, respectively. Without loss of generality, we assume $\mathbb{E}\{|s_k|^2\}=1$. $\boldsymbol{\rm h}_{{\rm r},k}\in\mathbb{C}^{N\times 1}$ and $\boldsymbol{\rm H}\in\mathbb{C}^{N\times M}$ denote the channels from the RIS to the $k$-th user and from the BS to the RIS, respectively.  $\boldsymbol{\rm \Theta}\triangleq{\textrm{diag}}(e^{j\theta_{1}},\cdots,e^{j\theta_{n}},\cdots e^{j\theta_{N}})$ is the diagonal phase-shift matrix, where $\theta_{n}$ denotes the corresponding phase shift. There exist two modes for each element, i.e., passive mode and off mode. The $n$-th element is switched to the passive mode when $\beta_n=1$ and is switched to the off mode when $\beta_n=0$. Define $\boldsymbol{\rm B}={\textrm{diag}}(\beta_{1},\cdots,\beta_{n},\cdots,\beta_{N})$ as the element on-off matrix for the passive RIS. $\boldsymbol{\rm n}\sim\mathcal{CN}(0,\delta^2\boldsymbol{\rm I}_M)$ denotes the additive white Gaussian noise (AWGN) at the BS.

To decode the signal from the $k$-th user, the BS applies a receive beamformer $\boldsymbol{\rm w}_k$ with $\|\boldsymbol{\rm w}_k\|^2=1$ to equalize the received signal for users
\begin{sequation}\label{eq2}
\begin{split}
y_k^{\rm pas}&=\boldsymbol{\rm w}_k^H\boldsymbol{\rm y}^{\rm pas}.
\end{split}
\end{sequation}
Then, the signal-to-interference-plus-noise ratio (SINR) of  the $k$-th user is given by
	\begin{sequation}\label{eq3}
	\begin{split}
	\gamma_k^{\rm pas}=\dfrac{p_k\left|\boldsymbol{\rm w}_k^H\boldsymbol{\rm H}^H\boldsymbol{\rm B}\boldsymbol{\rm \Theta}\boldsymbol{\rm h}_{{\rm r}, k}\right|^2}{\sum\limits_{i\not=k}^Kp_i\left|\boldsymbol{\rm w}_k^H\boldsymbol{\rm H}^H\boldsymbol{\rm B}\boldsymbol{\rm \Theta}\boldsymbol{\rm h}_{{\rm r}, i}\right|^2+\delta^2}.
 	\end{split}
	\end{sequation}
Thus, the achievable rate of the $k$-th user is formulated as
\begin{sequation}\label{eq4}
\begin{split}
R_k^{\rm pas}=\log_2(1+\gamma_k^{\rm pas}).
\end{split}
\end{sequation}
\subsection{Signal Transmission Model for Active RIS}
Different from the signal transmission model for the passive RIS, the active RIS will amplify noise, which cannot be ignored. Thus, the received signal at the BS is given by
\begin{sequation}\label{eq5}
\begin{split}
\boldsymbol{\rm y}^{\rm act}
&=\sum\limits_{k=1}^K\boldsymbol{\rm H}^H\boldsymbol{\rm A}\boldsymbol{\rm \Lambda}\boldsymbol{\rm \Theta}\boldsymbol{\rm h}_{{\rm r}, k}\sqrt{p_k}s_k + \boldsymbol{\rm H}^H\boldsymbol{\rm A}\boldsymbol{\rm \Lambda}\boldsymbol{\rm \Theta}\boldsymbol{\rm z} + \boldsymbol{\rm n},
\end{split}
\end{sequation}
where $\boldsymbol{\rm A}={\textrm{diag}}(\alpha_{1},\cdots,\alpha_{n},\cdots,\alpha_{N})$ denotes the element on-off matrix for the active RIS. The $n$-th reflecting element is switched to the active mode when $\alpha_n=1$ and is switched to the off mode when $\alpha_n=0$. $\boldsymbol{\rm \Lambda}={\textrm{diag}}(\rho_{1},\cdots,\rho_{n},\cdots,\rho_{N})$ denotes the reflecting amplification matrix, where $\rho_{n}> 1$ denotes the amplification factor of the $n$-th reflecting element. $\boldsymbol{\rm z}\in\mathbb{C}^{N\times 1}$ is the thermal noise introduced by the active reflecting elements due to signal amplification, which is assumed to follow the independent circularly symmetric complex Gaussian distribution, i.e., $\boldsymbol{\rm z}\sim\mathcal{CN}(\boldsymbol{0}, \sigma^2\boldsymbol{\rm I}_N)$. By applying the receive beamformer $\boldsymbol{\rm w}_k$ with $\|\boldsymbol{\rm w}_k\|^2=1$ at the BS, the SINR of the $k$-th user’s signal recovered is given by
		\begin{sequation}\label{eq6}
\begin{split}
\gamma_k^{\rm act}{=}\dfrac{p_k\left|\boldsymbol{\rm w}_k^H\boldsymbol{\rm H}^H\boldsymbol{\rm A}\boldsymbol{\rm \Lambda}\boldsymbol{\rm \Theta}\boldsymbol{\rm h}_{{\rm r}, k}\right|^2}{\sum\limits_{i\not=k}^Kp_i\left|\boldsymbol{\rm w}_k^H\boldsymbol{\rm H}^H\boldsymbol{\rm A}\boldsymbol{\rm \Lambda}\boldsymbol{\rm \Theta}\boldsymbol{\rm h}_{{\rm r}, i}\right|^2{+}\sigma^2\|\boldsymbol{\rm w}_{k}^H\boldsymbol{\rm H}^H\boldsymbol{\rm A}\boldsymbol{\rm \Lambda}\boldsymbol{\rm \Theta}\|^2+\delta^2}.
\end{split}
\end{sequation}
Then, the achievable rate of the $k$-th user for the active RIS is $R_k^{\rm act}=\log_2(1+\gamma_k^{\rm act})$.
\subsection{Power Model}
The power consumption at the active elements and the passive elements can be respectively expressed as
\begin{sequation}\label{eq7}
\begin{split}
P^{\rm pas}&=\sum\limits_{n=1}^N\beta_nP_{\textrm{C}},~~~~~P^{\rm act}=\sum\limits_{n=1}^N\alpha_{n}(P_{\textrm{C}}+P_{\textrm{DC}})+\sum\limits_{k=1}^Kp_k\|\boldsymbol{\rm A}\boldsymbol{\rm \Lambda}\boldsymbol{\rm \Theta}\boldsymbol{\rm h}_{{\rm r}, k}\|^2+\sigma^2\|\boldsymbol{\rm A}\boldsymbol{\rm \Lambda}\boldsymbol{\rm \Theta}\|_F^2,\\
\end{split}
\end{sequation}
where $P_{\textrm{C}}$ and $P_{\textrm{DC}}$ denote the power consumption of the circuit and the DC biasing power consumption, respectively. The total energy consumption of the system is given by
\begin{sequation}\label{eq8}
\begin{split}
P_{\rm Tot}^{\rm pas}&=\sum\limits_{k=1}^K(p_k+P_k)+P^{\rm BS}+P^{\rm pas},~~~~~P_{\rm Tot}^{\rm act}=\sum\limits_{k=1}^K(p_k+P_k)+P^{\rm BS}+P^{\rm act},
\end{split}
\end{sequation}
where $P^{\rm BS}$ and $P_k$ are the circuit power of the BS and each user, respectively. The amplification power of active elements for the active RIS is limited due to the total power budget at the active RIS. Then, the amplification power constraint is considered as follows
\begin{sequation}\label{eq9}
\begin{split}
& \sum\limits_{k=1}^Kp_k\|\boldsymbol{\rm A}\boldsymbol{\rm \Lambda}\boldsymbol{\rm \Theta}\boldsymbol{\rm h}_{{\rm r}, k}\|^2+\sigma^2\|\boldsymbol{\rm A}\boldsymbol{\rm \Lambda}\boldsymbol{\rm \Theta}\|_F^2\leq P_{\rm RIS}^{\max},
\end{split}
\end{sequation}
where $P_{\rm RIS}^{\max}$ is the maximum power threshold of the active RIS.
\section{The Element On-off  Algorithm for Passive RIS}
\subsection{Problem Formulation}
In this section, we aim to achieve the trade-off between the sum-rate and energy consumption for passive RIS-aided wireless communication networks. Mathematically, the energy efficiency maximization problem is expressed as follows
\begin{sequation}\label{eq10}
\begin{split}
\max\limits_{\mbox{\scriptsize$\begin{array}{c} 
		\boldsymbol{\rm \Theta},p_{k},\beta_n,\boldsymbol{\rm w}_k
		\end{array}$}} 
&\frac{\sum\limits_{k=1}^KR_k^{\rm pas}}{P_{\rm Tot}^{\rm pas}}\\
s.t.~{C_1}:& p_k\leq P_k^{\max},~~{C_2}:R_k^{\rm pas}\geq R_k^{\min},\\
{C_3}:&\beta_{n}\in\{0,1\},~~C_4:|[\boldsymbol{\rm B}\boldsymbol{\rm \Theta}]_{n,n}|=[\boldsymbol{\rm B}]_{n,n},
\end{split}
\end{sequation}
where $P_k^{\max}$ denotes the maximum transmit power of each user; $R_k^{\min}$ denotes the minimum transmission rate of the $k$-th user. Specifically, $C_1$ states that the maximum power of the $k$-th user; $C_2$ guarantees the minimum required transmission rate of each user; $C_3$ is element on-off factor constraint; $C_4$ is the unit-modulus constraints of passive elements. We note that problem (\ref{eq10}) is a highly non-convex optimization problem and challenging to obtain optimal solutions due to the fractional objective function and the unit-modulus constraint. In the next section, we develop a sub-optimal AO-based iterative algorithm to solve the problem (\ref{eq10}).

\subsection{Problem Transformation}
 It is worth noting that problem (\ref{eq10}) is a fractional programming problem. To this end, we utilize fraction programming methods proposed in \cite{i2} to decouple the variables in the problem (\ref{eq10}), which is given by the following Lemma.

\textbf{\textit{Lemma 1:}}  (Equivalent problem for energy efficiency maximization): By introducing auxiliary variables $t_k$, $r_k$, and $\eta$, the original problem in (\ref{eq10}) can be equivalently reformulated as follows
\begin{sequation}\label{eq12}
\begin{split}
\max\limits_{\mbox{\scriptsize$\begin{array}{c} 
		\boldsymbol{\rm \Theta},\beta_n\\
		t_k,r_k,p_k,
		\end{array}$}} 
EE_{\rm pas}&{=}\sum\limits_{k=1}^K\log_2(1+t_k)-\sum\limits_{k=1}^Kt_k+\sum\limits_{k=1}^Kf_{1,k}-\eta P_{\rm Tot}^{\rm pas}\\
s.t.~&C_1-C_4,
\end{split}
\end{sequation}
where
\begin{sequation}\label{eq13}
\begin{split}
f_{1,k}=\frac{(1+t_k)\gamma_k^{\rm pas}}{1+\gamma_k^{\rm pas}}&=2r_k\sqrt{(1+t_k)p_k|\boldsymbol{\rm w}_k^H\boldsymbol{\rm H}^H\boldsymbol{\rm B}\boldsymbol{\rm \Theta}\boldsymbol{\rm h}_{{\rm r}, k}|^2}-r_k^2(\sum_{i=1}^Kp_i|\boldsymbol{\rm w}_k^H\boldsymbol{\rm H}^H\boldsymbol{\rm B}\boldsymbol{\rm \Theta}\boldsymbol{\rm h}_{{\rm r}, i}|^2+\delta^2).
\end{split}
\end{sequation}
\begin{proof}
	We first introduce a  non-negative auxiliary variable $\eta$ and apply Dinkelbach's method to transform the energy efficiency problem to an equivalent parametrized non-fractional form. Then, we apply the method in \cite{i2} to deal with the transmission rate, please refer to \cite{i2} for the specific transformation process.
\end{proof}
After fixing other variables,  the optimal $t_k$ can be obtained by solving $\frac{\partial EE_{\rm pas}}{\partial t_k}=0$, i.e., 
\begin{sequation}\label{eq14}
\begin{split}
t_k=\gamma_k^{\rm pas}.
\end{split}
\end{sequation}
Meanwhile,  the optimal $r_k$ can be obtained by solving $\frac{\partial EE_{\rm pas}}{\partial r_k}=0$, i.e., 
\begin{sequation}\label{eq15}
\begin{split}
r_k=\frac{\sqrt{(1+t_k)p_k|\boldsymbol{\rm w}_k^H\boldsymbol{\rm H}^H\boldsymbol{\rm B}\boldsymbol{\rm \Theta}\boldsymbol{\rm h}_{{\rm r}, k}|^2}}{\sum\limits_{i=1}^Kp_i|\boldsymbol{\rm w}_k^H\boldsymbol{\rm H}^H\boldsymbol{\rm B}\boldsymbol{\rm \Theta}\boldsymbol{\rm h}_{{\rm r}, i}|^2+\delta^2}.
\end{split}
\end{sequation}

After fixing other variables,  define $\boldsymbol{\rm x}_k=\boldsymbol{\rm H}^H\boldsymbol{\rm B}\boldsymbol{\rm \Theta}\boldsymbol{\rm h}_{{\rm r}, k}$ and $\bar R_k^{\min}=2^{R_k^{\min}}-1$. Upon rearranging terms, the Lagrangian function can be written as (\ref{eq16}), where $\lambda_k$ and $\phi_k$ are corresponding non-negative Lagrangian multipliers.
\begin{figure*}\vspace{-20pt}
	\begin{sequation}\label{eq16}
		\begin{split}
		\mathcal{L}&=\sum\limits_{k=1}^K\log_2(1{+}t_k){-}\sum\limits_{k=1}^Kt_k{+}\sum\limits_{k=1}^K2r_k\sqrt{(1{+}t_k)p_k|\boldsymbol{\rm w}_k^H\boldsymbol{\rm x}_k|^2  }{-}\sum\limits_{k=1}^Kr_k^2(\sum\limits_{i=1}^Kp_i|\boldsymbol{\rm w}_k^H\boldsymbol{\rm x}_i|^2{+}\delta^2)\\
		&{-}\eta (\sum\limits_{k=1}^K(p_k{+}P_k){+}P^{\rm BS}{+}P^{\rm pas}){+}\sum\limits_{k=1}^K\lambda_k(P_k^{\max}{-}p_k){+}\sum\limits_{k=1}^K\phi_k\left(p_k|\boldsymbol{\rm w}_k^H\boldsymbol{\rm x}_k|^2{-}\bar R_k^{\min}(\sum\limits_{i\not=k}^Kp_i|\boldsymbol{\rm w}_k^H\boldsymbol{\rm x}_i|^2{+}\delta^2)\right).
		\end{split}
	\end{sequation}
	\hrulefill \vspace{-10pt}
\end{figure*}
Using the KKT conditions, $p_k$ is given by
\begin{sequation}\label{eq17}
\begin{split}
p_k=\frac{r_k^2(1+t_k)|\boldsymbol{\rm w}_k^H\boldsymbol{\rm x}_k|^2}{\left(\sum\limits_{i=1}^Kr_i^2|\boldsymbol{\rm w}_i^H\boldsymbol{\rm x}_k|^2+Y_k\right)^2},
\end{split}
\end{sequation}
where  and $Y_k=\eta+\lambda_k+\sum_{i\not=k}^K\phi_i\bar R_i^{\min}|\boldsymbol{\rm w}_i^H\boldsymbol{\rm x}_k|^2-\phi_k|\boldsymbol{\rm w}_k^H\boldsymbol{\rm x}_k|^2$. The optimal Lagrange multipliers $\lambda_k$ and $\phi_k$ can be obtained via sub-gradient methods.
\subsection{Receive Beamforming Vector Optimization} For given other variables, it is well-known that the linear MMSE detector is optimal receive beamforming to the problem (\ref{eq10}). Thus, the MMSE-based receive beamforming is written as
\begin{sequation}\label{eq11}
\begin{split}
\boldsymbol{\rm\bar w}_k=\left\{\sum\limits_{k=1}^Kp_k\boldsymbol{\rm h}_{1,k}\boldsymbol{\rm h}_{1,k}^H+\delta^2\boldsymbol{\rm I}_M\right\}^{-1}\sqrt{p_k}\boldsymbol{\rm h}_{1,k},
\end{split}
\end{sequation}
where $\boldsymbol{\rm h}_{1,k}=\boldsymbol{\rm H}^H\boldsymbol{\rm B}\boldsymbol{\rm \Theta}\boldsymbol{\rm h}_{{\rm r}, k}$. Then, we have $\boldsymbol{\rm w}_k=\frac{\boldsymbol{\rm\bar w}_k}{\|\boldsymbol{\rm\bar w}_k\|}$.

\subsection{Phase Shift and Element On-off Factor Optimization}
Next, we present the optimization of the phase shift $\boldsymbol{\rm \Theta}$ and the element on-off factor $\beta_n$ for other given variables. Then, problem (\ref{eq12}) can be reduced to
\begin{sequation}\label{eq18}
\begin{split}
&\max\limits_{\mbox{\scriptsize$\begin{array}{c} 
		\boldsymbol{\rm \Theta},\beta_n
		\end{array}$}} 
EE_{\rm pas}\\
s.t.~
&{C_2},{C_3},C_4.
\end{split}
\end{sequation}
To decouple the coupled variables $\boldsymbol{\rm B}$ and $\boldsymbol{\rm \Theta}$, we introduce a slack optimization variable $\boldsymbol{\rm v}=[v_1,\cdots,v_n,\cdots,v_N]^T\in \mathbb{C}^{N\times 1}$, where $v_n=\beta_ne^{j\theta_{n}}$. Then, we have the following transformation
\begin{sequation}\label{eq19}
\begin{split}
&|\boldsymbol{\rm w}_k^H\boldsymbol{\rm H}^H\boldsymbol{\rm B}\boldsymbol{\rm \Theta}\boldsymbol{\rm h}_{{\rm r}, k}|^2={\rm Tr}(\boldsymbol{\rm V}{\rm diag}(\boldsymbol{\rm w}_k^H\boldsymbol{\rm H}^H)\boldsymbol{\rm H}_{{\rm r},k}{\rm diag}(\boldsymbol{\rm w}_k^H\boldsymbol{\rm H}^H)^H),
 \end{split}
\end{sequation}
where $\boldsymbol{\rm H}_{{\rm r},k}=\boldsymbol{\rm h}_{{\rm r},k}\boldsymbol{\rm h}_{{\rm r},k}^H$ and $\boldsymbol{\rm V}=\boldsymbol{\rm v}\boldsymbol{\rm v}^H$.
Based on the above transformation, the objective function and $C_2$ can be rewritten as
\begin{sequation}\label{eq20}
\begin{split}
\overline{EE}_{\rm pas}&=\sum\limits_{k=1}^K\log_2(1+t_k)-\sum\limits_{k=1}^Kt_k+\sum\limits_{k=1}^K\bar f_{1,k}-\eta \bar P_{\rm Tot}^{\rm pas},
\end{split}
\end{sequation}
\begin{sequation}\label{eq21}
\begin{split}
&\bar C_2: p_k{\rm Tr}(\boldsymbol{\rm V}{\rm diag}(\boldsymbol{\rm w}_k^H\boldsymbol{\rm H}^H)\boldsymbol{\rm H}_{{\rm r},k}{\rm diag}(\boldsymbol{\rm w}_k^H\boldsymbol{\rm H}^H)^H)-\bar R_k^{\min}\times\\
&\left(\sum_{i\not=k}^Kp_i{\rm Tr}(\boldsymbol{\rm V}{\rm diag}(\boldsymbol{\rm w}_k^H\boldsymbol{\rm H}^H)\boldsymbol{\rm H}_{{\rm r},i}{\rm diag}(\boldsymbol{\rm w}_k^H\boldsymbol{\rm H}^H)^H)+\delta^2\right){\geq}0,
\end{split}
\end{sequation}
where
\begin{sequation}\label{eq22}
\begin{split}
\bar f_{1,k}&=2r_k\sqrt{(1+t_k)p_k{\rm Tr}(\boldsymbol{\rm V}{\rm diag}(\boldsymbol{\rm w}_k^H\boldsymbol{\rm H}^H)\boldsymbol{\rm H}_{{\rm r},k}{\rm diag}(\boldsymbol{\rm w}_k^H\boldsymbol{\rm H}^H)^H)}\\
&-r_k^2(\sum_{i=1}^Kp_i{\rm Tr}(\boldsymbol{\rm V}{\rm diag}(\boldsymbol{\rm w}_k^H\boldsymbol{\rm H}^H)\boldsymbol{\rm H}_{{\rm r},i}{\rm diag}(\boldsymbol{\rm w}_k^H\boldsymbol{\rm H}^H)^H)+\delta^2),
\end{split}
\end{sequation}
\begin{sequation}\label{eq23}
\begin{split}
\bar P_{\rm Tot}^{\rm pas}=\sum\limits_{k=1}^K(p_k+P_k)+P^{\rm BS}+\sum\limits_{n=1}^N|\boldsymbol{\rm v}_n|P_{\rm C}.
\end{split}
\end{sequation}
For the new introduced equation constraint $\boldsymbol{\rm V}=\boldsymbol{\rm v}\boldsymbol{\rm v}^H$, it can be equivalently rewritten as the following constraints\cite{i3}
\begin{sequation}\label{eq24}
C_{5a}:\left[                 
\begin{array}{cc}   
\boldsymbol{\rm V}& \boldsymbol{\rm v}\\  
\boldsymbol{\rm v}^H & 1\\  
\end{array}
\right] \succeq \boldsymbol{0},~ C_{5b}: {\rm Tr}(\boldsymbol{\rm V}-\boldsymbol{\rm v}\boldsymbol{\rm v}^H)\leq 0.      
\end{sequation}
Based on the first-order Taylor expansion, the lower bound of ${\rm Tr}(\boldsymbol{\rm v}\boldsymbol{\rm v}^H)$ can be derived as
\begin{sequation}\label{eq25}
\begin{split}
{\rm Tr}(\boldsymbol{\rm v}\boldsymbol{\rm v}^H)\geq -\|\boldsymbol{\rm\bar v}\|^2+2{\rm Tr}(\boldsymbol{\rm\bar v}^H\boldsymbol{\rm v}),
\end{split}
\end{sequation}
where $\boldsymbol{\rm\bar v}$ is the previous iteration of $\boldsymbol{\rm v}$. Thus, by substituting the lower
bound in (\ref{eq25}) into $C_{5b}$, $C_{5b}$ can be rewritten as
\begin{sequation}\label{eq26}
\begin{split}
\bar C_{5b}:{\rm Tr}(\boldsymbol{\rm V})\leq -\|\boldsymbol{\rm\bar v}\|^2+2{\rm Tr}(\boldsymbol{\rm\bar v}^H\boldsymbol{\rm v}).
\end{split}
\end{sequation}
Then, problem (\ref{eq18}) can be transformed into the following convex problem
\begin{sequation}\label{eq27}
\begin{split}
&\max\limits_{\mbox{\scriptsize$\begin{array}{c} 
		\boldsymbol{\rm V},\boldsymbol{\rm v}
		\end{array}$}} 
\overline{EE}_{\rm pas}\\
s.t.~&\bar C_4:|\boldsymbol{\rm v}_n|\in\{0,1\}, {\bar  C_2},C_{5a},\bar C_{5b}.
\end{split}
\end{sequation}
It is noted that the problem (\ref{eq27}) is still a non-convex problem due to the binary variable $|\boldsymbol{\rm v}_n|$ in $\bar C_4$. Then, we can transform $\bar C_4$ as
\begin{sequation}\label{eq28}
\begin{split}
C_{4a}:  |\boldsymbol{\rm v}_n|\leq 1, C_{4b}: |\boldsymbol{\rm v}_n|-|\boldsymbol{\rm v}_n|^2\leq 0.
\end{split}
\end{sequation}
However, $C_{4b}$ is still a non-convex constraint. Thus,  the non-convex constraint $C_{4b}$ can be  linearized as
\begin{sequation}\label{eq30}
\begin{split}
\bar C_{4b}: |\boldsymbol{\rm v}_n|+|\boldsymbol{\rm\bar v}_n|^2-2{\rm Re}(\boldsymbol{\rm v}_n^*\boldsymbol{\rm\bar v}_n) \leq 0,
\end{split}
\end{sequation}

Finally, problem (\ref{eq27}) is transformed into the following convex optimization problem:
\begin{sequation}\label{eq31}
\begin{split}
&\max\limits_{\mbox{\scriptsize$\begin{array}{c} 
		\boldsymbol{\rm V},\boldsymbol{\rm v}
		\end{array}$}} 
\overline{EE}_{\rm pas}\\
s.t.~& {\bar  C_2},C_{4a},\bar C_{4b}, C_{5a},\bar C_{5b}.
\end{split}
\end{sequation}
Problem (\ref{eq31}) is a convex optimization problem and can be solved directly by the standard convex optimization techniques. 

\section{The Element On-off  Algorithm for Active RIS}
The traditional passive RIS suffers from the ``double fading'' effect, which has become a major bottleneck in restricting the performance of RIS-aided communications. Thus,  we investigate the impact of the element on-off mechanism on active RIS-aided communications in this section. Different from the passive RIS, the optimization process for the active RIS is more complex since it involves optimizing not only the phase shift but also the amplification factor. Furthermore, the active RIS has higher power consumption and is subject to the maximum power constraint of the active RIS. This means that the element on-off algorithm for the passive RIS cannot be directly applied to the active RIS, thus a new  element on-off algorithm needs to be designed.

\subsection{Problem Formulation}
Based on the system model for the active RIS, the energy efficiency maximization problem for the active RIS is formulated as follows
\begin{sequation}\label{eq32}
\begin{split}
&\max\limits_{\mbox{\scriptsize$\begin{array}{c} 
		\boldsymbol{\rm \Theta},\alpha_{n}\\
		\boldsymbol{\rm w}_k,\rho_{n},p_k
		\end{array}$}} 
\frac{\sum\limits_{k=1}^KR_k^{\rm act}}{P_{\rm Tot}^{\rm act}}\\
s.t.~{C_1}:& p_k\leq P_k^{\max},~~~~{C_2}: \sum\limits_{k=1}^Kp_k\|\boldsymbol{\rm A}\boldsymbol{\rm \Lambda}\boldsymbol{\rm \Theta}\boldsymbol{\rm h}_{{\rm r}, k}\|^2+\sigma^2\|\boldsymbol{\rm A}\boldsymbol{\rm \Lambda}\boldsymbol{\rm \Theta}\|_F^2\leq P_{\rm RIS}^{\max},\\
{C_3}:&R_k^{\rm act}\geq R_k^{\min},~~{C_4}:\alpha_{n}\in\{0,1\},\\
\end{split}
\end{sequation}
where $C_2$ states that the maximum power of the active elements should not exceed $P_{\rm RIS}^{\max}$. Compared to the passive RIS, although the active RIS avoids the unit-modulus constraint, it also introduces the additional non-convex constraint $C_2$, which aggravates the coupling between the optimization variables. Next, we propose another AO-based algorithm to solve the problem (\ref{eq32}).
\subsection{Problem Transformation}
 Based on Lemma 1, the original problem in (\ref{eq32}) can be equivalently reformulated as follows
\begin{sequation}\label{eq34}
\begin{split}
\max\limits_{\mbox{\scriptsize$\begin{array}{c} 
		\boldsymbol{\rm \Theta},\alpha_{n},\beta_n\\
		t_k,r_k,\rho_{n},
		\end{array}$}} 
EE_{\rm act}&=\sum\limits_{k=1}^K\log_2(1+t_k)-\sum\limits_{k=1}^Kt_k+\sum\limits_{k=1}^Kf_{2,k}-\eta P_{\rm Tot}^{\rm act}\\
s.t.~&C_1-C_5,
\end{split}
\end{sequation}
where
\begin{sequation}\label{eq35}
\begin{split}
f_{2,k}=&2r_k\sqrt{(1+t_k)p_k|\boldsymbol{\rm w}_k^H\boldsymbol{\rm H}^H\boldsymbol{\rm A}\boldsymbol{\rm \Lambda}\boldsymbol{\rm \Theta}\boldsymbol{\rm h}_{{\rm r}, k}|^2}-r_k^2(\sum_{i=1}^Kp_i|\boldsymbol{\rm w}_k^H\boldsymbol{\rm H}^H\boldsymbol{\rm A}\boldsymbol{\rm \Lambda}\boldsymbol{\rm \Theta}\boldsymbol{\rm h}_{{\rm r}, i}|^2{+}\sigma^2\|\boldsymbol{\rm w}_{k}^H\boldsymbol{\rm H}^H\boldsymbol{\rm A}\boldsymbol{\rm \Lambda}\boldsymbol{\rm \Theta}\|^2{+}\delta^2).
\end{split}
\end{sequation}
For maximizing energy efficiency $EE_{\rm act}$ iteratively over $t_k$ and $r_k$, we find closed-form update equations as
\begin{sequation}\label{eq36}
\begin{split}
t_k=\gamma_k^{\rm act},
\end{split}
\end{sequation}
\begin{sequation}\label{eq37}
\begin{split}
r_k=\frac{\sqrt{p_k(1+t_k)|\boldsymbol{\rm w}_k^H\boldsymbol{\rm H}^H\boldsymbol{\rm A}\boldsymbol{\rm \Lambda}\boldsymbol{\rm \Theta}\boldsymbol{\rm h}_{{\rm r}, k}|^2}}{\sum\limits_{i=1}^Kp_i|\boldsymbol{\rm w}_k^H\boldsymbol{\rm H}^H\boldsymbol{\rm A}\boldsymbol{\rm \Lambda}\boldsymbol{\rm \Theta}\boldsymbol{\rm h}_{{\rm r}, i}|^2{+}\sigma^2\|\boldsymbol{\rm w}_{k}^H\boldsymbol{\rm H}^H\boldsymbol{\rm A}\boldsymbol{\rm \Lambda}\boldsymbol{\rm \Theta}\|^2{+}\delta^2}.
\end{split}
\end{sequation}
After fixing other variables,  define $\boldsymbol{\rm x}_k=\boldsymbol{\rm H}^H\boldsymbol{\rm A}\boldsymbol{\rm \Lambda}\boldsymbol{\rm \Theta}\boldsymbol{\rm h}_{{\rm r}, k}$, the Lagrangian function can be written as (\ref{eq38}), where $\lambda_k$, $\phi_k$, and $\kappa$ are corresponding non-negative Lagrangian multipliers.
\begin{figure*}\vspace{-20pt}
	\begin{sequation}\label{eq38}
	\begin{split}
	&\mathcal{F}=\sum\limits_{k=1}^K\log_2(1{+}t_k){-}\sum\limits_{k=1}^Kt_k{+}\sum\limits_{k=1}^K2r_k\sqrt{(1{+}t_k)p_k|\boldsymbol{\rm w}_k^H\boldsymbol{\rm x}_k|^2  }{-}\sum\limits_{k=1}^Kr_k^2(\sum\limits_{i=1}^Kp_i|\boldsymbol{\rm w}_k^H\boldsymbol{\rm x}_i|^2{+}\sigma^2\|\boldsymbol{\rm w}_{k}^H\boldsymbol{\rm H}^H\boldsymbol{\rm A}\boldsymbol{\rm \Lambda}\boldsymbol{\rm \Theta}\|^2{+}\delta^2)\\	
	&{-}\eta (\sum\limits_{k=1}^K(p_k+P_k){+}P^{\rm BS}{+}\sum\limits_{n=1}^N\alpha_{n}(P_{\textrm{C}}{+}P_{\textrm{DC}}){+}\sum\limits_{k=1}^Kp_k\|\boldsymbol{\rm A}\boldsymbol{\rm \Lambda}\boldsymbol{\rm \Theta}\boldsymbol{\rm h}_{{\rm r}, k}\|^2{+}\sigma^2\|\boldsymbol{\rm A}\boldsymbol{\rm \Lambda}\boldsymbol{\rm \Theta}\|_F^2){+}\sum\limits_{k=1}^K\lambda_k(P_k^{\max}{-}p_k){+}\kappa\left\{P_{\rm RIS}^{\max}\right.\\
	&\left.{-}(\sum\limits_{k=1}^Kp_k\|\boldsymbol{\rm A}\boldsymbol{\rm \Lambda}\boldsymbol{\rm \Theta}\boldsymbol{\rm h}_{{\rm r}, k}\|^2{+}\sigma^2\|\boldsymbol{\rm A}\boldsymbol{\rm \Lambda}\boldsymbol{\rm \Theta}\|_F^2)\right\}{+}\sum\limits_{k=1}^K\phi_k\left\{p_k|\boldsymbol{\rm w}_k^H\boldsymbol{\rm x}_k|^2{-}\bar R_k^{\min}(\sum\limits_{i\not=k}^Kp_i|\boldsymbol{\rm w}_k^H\boldsymbol{\rm x}_i|^2){+}\sigma^2\|\boldsymbol{\rm w}_{k}^H\boldsymbol{\rm H}^H\boldsymbol{\rm A}\boldsymbol{\rm \Lambda}\boldsymbol{\rm \Theta}\|^2{+}\delta^2\right\}.
	\end{split}
	\end{sequation}
	\hrulefill 
\end{figure*}
Using KKT conditions, $p_k$ is given by
\begin{sequation}\label{eq39}
\begin{split}
p_k=\frac{r_k^2(1+t_k)|\boldsymbol{\rm w}_k^H\boldsymbol{\rm x}_k|^2}{\left(\sum\limits_{i=1}^Kr_i^2|\boldsymbol{\rm w}_i^H\boldsymbol{\rm x}_k|^2+(\eta+\kappa)\|\boldsymbol{\rm A}\boldsymbol{\rm \Lambda}\boldsymbol{\rm \Theta}\boldsymbol{\rm h}_{{\rm r}, k}\|^2+Y_k\right)^2},
\end{split}
\end{sequation}
where  and $Y_k=\eta+\lambda_k+\sum_{i\not=k}^K\phi_i\bar R_i^{\min}|\boldsymbol{\rm w}_i^H\boldsymbol{\rm x}_k|^2-\phi_k|\boldsymbol{\rm w}_k^H\boldsymbol{\rm x}_k|^2$. The optimal Lagrange multipliers $\lambda_k$, $\kappa$, and $\phi_k$ can be obtained via sub-gradient methods.
\subsection{Receive Beamforming Vector Optimization}
For given other variables,  based on the linear MMSE principle, the receive beamforming for the active RIS is written as
\begin{sequation}\label{eq33}
\begin{split}
\boldsymbol{\rm\bar w}_k&=\left\{\sum\limits_{k=1}^Kp_k\boldsymbol{\rm h}_{2,k}\boldsymbol{\rm h}_{2,k}^H+\sigma^2\boldsymbol{\rm H}^H\boldsymbol{\rm A}\boldsymbol{\rm \Lambda}\boldsymbol{\rm \Theta}\boldsymbol{\rm \Theta}^H\boldsymbol{\rm \Lambda}^H\boldsymbol{\rm A}^H\boldsymbol{\rm H}+\delta^2\boldsymbol{\rm I}_M\right\}^{-1}\!\sqrt{p_k}\boldsymbol{\rm h}_{1,k},
\end{split}
\end{sequation}
where $\boldsymbol{\rm h}_{2,k}=\boldsymbol{\rm H}^H\boldsymbol{\rm A}\boldsymbol{\rm \Lambda}\boldsymbol{\rm \Theta}\boldsymbol{\rm h}_{{\rm r}, k}$. Then, we have $\boldsymbol{\rm w}_k=\frac{\boldsymbol{\rm\bar w}_k}{\|\boldsymbol{\rm\bar w}_k\|}$.
\subsection{Amplification Factor, Phase Shift, and Element On-off Factor Optimization}
In this subsection, we present the joint optimization of the amplification factor, phase shift, and element on-off factor by fixing other variables. In particular,  problem (\ref{eq34}) can be reduced to
\begin{sequation}\label{eq40}
\begin{split}
\max\limits_{\mbox{\scriptsize$\begin{array}{c} 
		\rho_{n},\boldsymbol{\rm \Theta},\alpha_n
		\end{array}$}} 
~~&EE_{\rm act}\\
s.t.~&{C_2},{ C_3},C_4.
\end{split}
\end{sequation}
It’s worth noting that matrices $\boldsymbol{\rm \Lambda}$ and $\boldsymbol{\rm \Theta}$ are always coupled in the product form. Hence, we rewrite the product term $\boldsymbol{\rm \Lambda}\boldsymbol{\rm \Theta}$ as $\boldsymbol{\rm \Phi}$, where $\boldsymbol{\rm \Phi}={\rm diag}(\boldsymbol{{\rm \phi}})$ and $\boldsymbol{{\rm \phi}}=[\phi_1,\cdots,\phi_n,\cdots,\phi_N]^T$. 
However, $\boldsymbol{\rm A}\boldsymbol{\rm \Phi}$ is still coupled. To decouple the coupled variables $\boldsymbol{\rm A}$ and $\boldsymbol{\rm \Phi}$, we introduce a slack optimization variable $\boldsymbol{\rm u}=[u_1,\cdots,u_n,\cdots,u_N]^T\in \mathbb{C}^{N\times 1}$  and ${\rm diag}(\boldsymbol{\rm u}^H)=\boldsymbol{\rm A}\boldsymbol{\rm \Phi}$. By adopting the Big-M formulation \cite{i4}, ${\rm diag}(\boldsymbol{\rm u}^H)=\boldsymbol{\rm A}\boldsymbol{\rm \Phi}$ can be converted into a set of equivalent constraints as follows:
\begin{sequation}\label{eq41}
\begin{split}
&C_{5a}:\phi_{n}^*-(1-\alpha_{n})M\leq u_{n},~~C_{5b}:u_{n}\leq \phi_{n}^*+(1-\alpha_{n})M,\\
&C_{5c}:-\alpha_{n}M\leq u_{n},~~~~~~~~~~~~~~C_{5d}:u_{n}\leq \alpha_{n}M,
\end{split}
\end{sequation}
where $u_{n}$ obtains the same value as $\phi_{n}^*$ by $C_{5a}$ and $C_{5b}$ when $\alpha_{n}=1$; $u_{n}$ is forced to zero by constraints $C_{5c}$ and $C_{5d}$ when $\alpha_{n}=0$.  $M$ is an arbitrarily large constant. Then, we introduce a new variable $\boldsymbol{\rm U}=\boldsymbol{\rm u}\boldsymbol{\rm u}^H$. For the new introduced equation constraint $\boldsymbol{\rm U}=\boldsymbol{\rm u}\boldsymbol{\rm u}^H$, it can be equivalently rewritten as the following constraints
\begin{sequation}\label{eq42}
C_{6a}:\left[                 
\begin{array}{cc}   
\boldsymbol{\rm U}& \boldsymbol{\rm u}\\  
\boldsymbol{\rm u}^H & 1\\  
\end{array}
\right] \succeq \boldsymbol{0},~ C_{6b}: {\rm Tr}(\boldsymbol{\rm U}-\boldsymbol{\rm u}\boldsymbol{\rm u}^H)\leq 0.      
\end{sequation}
Based on the first-order Taylor expansion, the lower bound of ${\rm Tr}(\boldsymbol{\rm u}\boldsymbol{\rm u}^H)$ can be derived as
\begin{sequation}\label{eq43}
\begin{split}
{\rm Tr}(\boldsymbol{\rm u}\boldsymbol{\rm u}^H)\geq -\|\boldsymbol{\rm\bar u}\|^2+2{\rm Tr}(\boldsymbol{\rm\bar u}^H\boldsymbol{\rm u}),
\end{split}
\end{sequation}
where $\boldsymbol{\rm\bar u}$ is the previous iteration of $\boldsymbol{\rm u}$. Thus, by substituting the lower bound in (\ref{eq43}) into $C_{6b}$, $C_{6b}$ can be rewritten as
\begin{sequation}\label{eq44}
\begin{split}
\bar C_{6b}:{\rm Tr}(\boldsymbol{\rm U})\leq -\|\boldsymbol{\rm\bar u}\|^2+2{\rm Tr}(\boldsymbol{\rm\bar u}^H\boldsymbol{\rm u}).
\end{split}
\end{sequation}

Then, based on the introduced variable, we have the following equivalent transformations
\begin{sequation}\label{eq45}
\begin{split}
|\boldsymbol{\rm w}_k^H\boldsymbol{\rm H}^H\boldsymbol{\rm A}\boldsymbol{\rm \Phi}\boldsymbol{\rm h}_{{\rm r}, k}|^2&={\rm Tr}(\boldsymbol{\rm U}{\rm diag}(\boldsymbol{\rm w}_k^H\boldsymbol{\rm H}^H)\boldsymbol{\rm H}_{{\rm r},k}{\rm diag}(\boldsymbol{\rm H}\boldsymbol{\rm w}_k)),\\
\|\boldsymbol{\rm w}_k^H\boldsymbol{\rm H}^H\boldsymbol{\rm A}\boldsymbol{\rm \Phi}\|^2&={\rm Tr}({\rm diag}(\boldsymbol{\rm H}\boldsymbol{\rm w}_k)\boldsymbol{\rm U}{\rm diag}(\boldsymbol{\rm w}_k^H\boldsymbol{\rm H}^H)),\\
\|\boldsymbol{\rm A}\boldsymbol{\rm \Phi}\boldsymbol{\rm h}_{{\rm r}, k}\|^2&={\rm Tr}({\rm diag}(\boldsymbol{\rm h}_{{\rm r},k}^H)\boldsymbol{\rm U}{\rm diag}(\boldsymbol{\rm h}_{{\rm r},k})),\\
\|\boldsymbol{\rm A}\boldsymbol{\rm \Phi}\|_F^2&={\rm Tr}(\boldsymbol{\rm U}).
\end{split}
\end{sequation}

Based on the above transformations, the objective function, $C_2$, and $C_3$ can be rewritten as
\begin{sequation}\label{eq46}
\begin{split}
\overline{EE}_{\rm act}=\sum\limits_{k=1}^K\log_2(1+t_k)-\sum\limits_{k=1}^Kt_k+\sum\limits_{k=1}^K\bar f_{2,k}-\eta \bar P_{\rm Tot}^{\rm act},
\end{split}
\end{sequation}
\begin{sequation}\label{eq47}
\begin{split}
{\bar C_2}: \sum\limits_{k=1}^Kp_k{\rm Tr}({\rm diag}(\boldsymbol{\rm h}_{{\rm r},k}^H)\boldsymbol{\rm U}{\rm diag}(\boldsymbol{\rm h}_{{\rm r},k}))+\sigma^2{\rm Tr}(\boldsymbol{\rm U})\leq P_{\rm RIS}^{\max},
\end{split}
\end{sequation}
\begin{sequation}\label{eq48}
\begin{split}
\bar C_3:&p_k{\rm Tr}(\boldsymbol{\rm U}{\rm diag}(\boldsymbol{\rm w}_k^H\boldsymbol{\rm H}^H)\boldsymbol{\rm H}_{{\rm r},k}{\rm diag}(\boldsymbol{\rm H}\boldsymbol{\rm w}_k))-\bar R_k^{\min}\left(\sum_{i\not=k}^Kp_i{\rm Tr}(\boldsymbol{\rm U}{\rm diag}(\boldsymbol{\rm w}_k^H\boldsymbol{\rm H}^H)\boldsymbol{\rm H}_{{\rm r},i}{\rm diag}(\boldsymbol{\rm H}\boldsymbol{\rm w}_k))\right.\\
&\left.+\sigma^2{\rm Tr}({\rm diag}(\boldsymbol{\rm H}\boldsymbol{\rm w}_k)\boldsymbol{\rm U}{\rm diag}(\boldsymbol{\rm w}_k^H\boldsymbol{\rm H}^H))+\delta^2\right)\geq 0,
\end{split}
\end{sequation}
where
\begin{sequation}\label{eq49}
\begin{split}
&\bar f_{2,k}=2r_k\sqrt{(1+t_k)p_k{\rm Tr}(\boldsymbol{\rm U}{\rm diag}(\boldsymbol{\rm w}_k^H\boldsymbol{\rm H}^H)\boldsymbol{\rm H}_{{\rm r},k}{\rm diag}(\boldsymbol{\rm H}\boldsymbol{\rm w}_k))}\\
&-r_k^2\left(\sum_{i=1}^Kp_i{\rm Tr}(\boldsymbol{\rm U}{\rm diag}(\boldsymbol{\rm w}_k^H\boldsymbol{\rm H}^H)\boldsymbol{\rm H}_{{\rm r},i}{\rm diag}(\boldsymbol{\rm H}\boldsymbol{\rm w}_k))+\sigma^2{\rm Tr}({\rm diag}(\boldsymbol{\rm H}\boldsymbol{\rm w}_k)\boldsymbol{\rm U}{\rm diag}(\boldsymbol{\rm w}_k^H\boldsymbol{\rm H}^H))+\delta^2\right),
\end{split}
\end{sequation}
\begin{sequation}\label{eq50}
\begin{split}
\bar P_{\rm Tot}^{\rm act}&=\sum\limits_{k=1}^K(p_k+P_k)+P^{\rm BS}+\sum\limits_{n=1}^N\alpha_{n}(P_{\textrm{C}}+P_{\textrm{DC}})+\sum\limits_{k=1}^Kp_k{\rm Tr}({\rm diag}(\boldsymbol{\rm h}_{{\rm r},k}^H)\boldsymbol{\rm U}{\rm diag}(\boldsymbol{\rm h}_{{\rm r},k}))+\sigma^2{\rm Tr}(\boldsymbol{\rm U}).
\end{split}
\end{sequation}

 Since $\alpha_{n}$ is a binary variable, we equivalently transform $C_4$ as
\begin{sequation}\label{eq51}
\begin{split}
C_{7a}:\alpha_{n}-\alpha_{n}^2\leq 0,~~~ C_{7b}:0\leq \alpha_{n}\leq 1.
\end{split}
\end{sequation}
Furthermore, based on the first-order Taylor expansion, the lower bound of $\alpha_{n}^2$ can be derived as $\alpha_{n}^2\geq\bar\alpha_{n}^2 + 2\bar\alpha_{n}(\alpha_{n}-\bar\alpha_{n})$. Then, $C_{7a}$ can be rewritten as
\begin{sequation}\label{eq52}
\begin{split}
\bar C_{7a}:\alpha_{n}-(\bar\alpha_{n}^2 + 2\bar\alpha_{n}(\alpha_{n}-\bar\alpha_{n}))\leq 0,
\end{split}
\end{sequation}
where $\bar\alpha_{k,n}$ is the previous iteration of $\alpha_{k,n}$.
Then, problem (\ref{eq40}) can be transformed into
\begin{sequation}\label{eq53}
\begin{split}
\max\limits_{\mbox{\scriptsize$\begin{array}{c} 
		\boldsymbol{\rm \Phi},\boldsymbol{\rm U},\boldsymbol{\rm u},\alpha_n
		\end{array}$}} 
~~&\overline{EE}_{\rm act}\\
s.t.~&{\bar C_2},{\bar C_3},\bar C_{5a}-C_{5d}, C_{6a}, \bar C_{6b},\bar C_{7a}, C_{7b}.
\end{split}
\end{sequation}
Problem (\ref{eq53}) is a convex optimization problem and can be solved directly by the standard convex optimization techniques. 
\section{Number Configuration of Reflecting Elements}
In this section, we investigate the number configuration for the passive RIS and the active RIS to gain more insight, where the same total power budget, i.e., $P_{\rm Tol}^{\rm act}=P_{\rm Tol}^{\rm pas}=P_{\rm Tol}$ is considered for fair comparison. To facilitate our analysis, we consider a single-user communication scenario, where the BS is equipped with a single antenna, i.e., $\boldsymbol{\rm H}\in \mathbb{C}^{N\times M}\rightarrow\boldsymbol{\rm h}\in \mathbb{C}^{N\times 1}$, the user performs the uplink information transmission at a maximum power $P^{\max}$, the amplification factor of each reflecting element is the same, i.e., $\rho_{n}=\rho,\forall n$ \cite{d1}. Before getting into the details of this section, we would like to highlight that the number configuration in this paper is different from previous works (see, e.g., \cite{d10} and \cite{i4-1}) in that the former aims at optimizing the number of active (or equivalently, passive) elements given a fixed number of the total number of reflecting elements while the latter investigates on how to minimize the total number of reflecting elements.

\subsection{Number Configuration for Passive RIS}
In this subsection, we focus on the number configuration for the passive RIS. The energy efficiency maximization problem can be transformed into
\begin{sequation}\label{eq54}
\begin{split}
\max\limits_{\mbox{\scriptsize$\begin{array}{c} 
		\boldsymbol{\rm \Theta},\beta_n
		\end{array}$}} 
&\frac{\log_2(1+\frac{P^{\max}\left|\boldsymbol{\rm h}^H\boldsymbol{\rm B}\boldsymbol{\rm \Theta}\boldsymbol{\rm h}_{{\rm r}}\right|^2}{\delta^2})}{P_{\rm Tot}}\\
s.t.~{C_1}:&\log_2(1+\frac{P^{\max}\left|\boldsymbol{\rm h}^H\boldsymbol{\rm B}\boldsymbol{\rm \Theta}\boldsymbol{\rm h}_{{\rm r}}\right|^2}{\delta^2})\geq R^{\min},~~{C_2}:\beta_{n}\in\{0,1\},~~C_3:|[\boldsymbol{\rm B}\boldsymbol{\rm \Theta}]_{n,n}|=[\boldsymbol{\rm B}]_{n,n}.
\end{split}
\end{sequation}

With the given total power budget, the optimization problem (\ref{eq54}) can be transformed into the following problem, i.e.,
\begin{sequation}\label{eq55}
\begin{split}
&\max\limits_{\mbox{\scriptsize$\begin{array}{c} 
		\boldsymbol{\rm \Theta},\beta_n
		\end{array}$}} 
\log_2(1+\frac{P^{\max}\left|\boldsymbol{\rm h}^H\boldsymbol{\rm B}\boldsymbol{\rm \Theta}\boldsymbol{\rm h}_{{\rm r}}\right|^2}{\delta^2})\\
s.t.~&{C_1}-C_3,C_4:\sum\limits_{n=1}^N\beta_nP_{\rm C}\leq P_{\rm Tot}-P^{\max}-P-P^{\rm BS}\triangleq C.
\end{split}
\end{sequation}
\subsubsection{Phase-shift Optimization}Given the element on-off factor, it can be easily shown that the transmission rate in (\ref{eq55}) is maximized when the optimal phase-shift matrix of the RIS aligns the cascaded user-RIS-BS channel\cite{i5}, i.e.,
\begin{sequation}\label{eq56}
\begin{split}
\theta_{n}=\arg([\boldsymbol{\rm h}]_n)-\arg([\boldsymbol{\rm h}_{\rm r}]_n).
\end{split}
\end{sequation}
Then, by substituting the optimal phase shift in (\ref{eq56}) into the transmission rate expression, we have
\begin{sequation}\label{eq57}
\begin{split}
&\frac{P^{\max}\left|\boldsymbol{\rm h}^H\boldsymbol{\rm B}\boldsymbol{\rm \Theta}\boldsymbol{\rm h}_{{\rm r}}\right|^2}{\delta^2}\overset{(a)}{=}\frac{P^{\max}\left|\sum\limits_{n=1}^N\beta_{n}|h_n||h_{{\rm r},n}|\right|^2}{\delta^2}\overset{(b)}{\geq} \frac{P^{\max}|h|^2|h_{{\rm r}}|^2(\sum\limits_{n=1}^N\beta_{n})^2}{\delta^2},
\end{split}
\end{sequation}
where $(a)$ utilizes the optimal design of $\boldsymbol{\rm \Theta}$, $(b)$ utilizes $|h|=\min\{|h_n|\}$ and $|h_{{\rm r}}|=\min\{|h_{{\rm r},n}|\}$. $h_n$ and $h_{{\rm r},n}$ are the $n$-th elements of $\boldsymbol{\rm h}$ and $\boldsymbol{\rm h}_{{\rm r}}$.
\subsubsection{Number Configuration}
Next, we optimize the number of reflecting elements in the passive mode. Define the number of reflecting elements in the passive mode as $N_{\rm pas}=\sum_{n=1}^N\beta_n$, then the number of reflecting elements in the off mode is $N-N_{\rm pas}$. Furthermore, since the logarithm function is a monotonic increasing function, the problem (\ref{eq55}) can be reduced to
\begin{sequation}\label{eq58}
\begin{split}
\max\limits_{\mbox{\scriptsize$\begin{array}{c} 
		N_{\rm pas}
		\end{array}$}} 
&\frac{P^{\max}|h|^2|h_{{\rm r}}|^2N_{\rm pas}^2}{\delta^2}\\
{\bar C_1}:&\frac{P^{\max}|h|^2|h_{{\rm r}}|^2N_{\rm pas}^2}{\delta^2}\geq \bar R^{\min},~~{\bar C_4}: N_{\rm pas}\leq \left \lfloor \frac{C}{P_{\rm C}} \right \rfloor.
\end{split}
\end{sequation}
By analyzing problem (\ref{eq58}), we can obtain the lower bound and upper bound of $N_{\rm pas}$
\begin{sequation}\label{eq59}
\begin{split}
N_{\rm pas}^{\rm L}=\left \lceil \sqrt{\frac{\delta^2\bar R^{\min}}{P^{\max}|h|^2|h_{\rm r}|^2}} \right \rceil,
\end{split}
\end{sequation}
\begin{sequation}\label{eq60}
\begin{split}
N_{\rm pas}^{\rm U}=\min \left\{\left \lfloor \frac{C}{P_{\rm C}} \right \rfloor, N\right\}.
\end{split}
\end{sequation}
We can observe that the objective function is a monotonically increasing function of $N_{\rm pas}$. Then, the optimal closed-form solution of $N_{\rm pas}$ is given by
\begin{sequation}\label{eq61}
N_{\rm pas}{=}\left\{
\begin{split}
&N_{\rm pas}^{\rm U},~~~~~~~~~~~~N_{\rm pas}^{\rm L}\leq N_{\rm pas}^{\rm U},\\
 &\textrm{Infeasible},~~~~~~~\textrm{Otherwise}.
\end{split}
\right.
\end{sequation}
\textbf{\textit{ Remark 1:}} From (\ref{eq61}), given the same total power budget, since the passive RIS does not exist the amplified thermal noise introduced by the active RIS, the number of reflecting elements in the passive mode will take the maximum value, but will be limited by the total power budget. When the total power budget is small or the circuit power consumption of  the reflecting elements is large, it may cause a communication outage. In addition, the increasing number of reflecting elements in the passive mode can also effectively alleviate the power pressure of users.

\subsection{Number Configuration for Active RIS}
In this subsection, we focus  on the number configuration for the active RIS. The energy efficiency maximization problem can be transformed into
\begin{sequation}\label{eq62}
\begin{split}
\max\limits_{\mbox{\scriptsize$\begin{array}{c} 
		\boldsymbol{\rm \Theta},\rho,\beta_n
		\end{array}$}} 
&\dfrac{\log_2(1+\frac{P^{\max}\left|\boldsymbol{\rm h}^H\boldsymbol{\rm A}\boldsymbol{\rm \Lambda}\boldsymbol{\rm \Theta}\boldsymbol{\rm h}_{{\rm r}}\right|^2}{\sigma^2\|\boldsymbol{\rm h}^H\boldsymbol{\rm A}\boldsymbol{\rm \Lambda}\boldsymbol{\rm \Theta}\|^2+\delta^2})}{P_{\rm Tot}}\\
s.t.~{C_1}:& P^{\max}\|\boldsymbol{\rm A}\boldsymbol{\rm \Lambda}\boldsymbol{\rm \Theta}\boldsymbol{\rm h}_{{\rm r}}\|^2+\sigma^2\|\boldsymbol{\rm A}\boldsymbol{\rm \Lambda}\boldsymbol{\rm \Theta}\|_F^2\leq P_{\rm RIS}^{\max},\\
C_2:&\log_2(1+\frac{P^{\max}\left|\boldsymbol{\rm h}^H\boldsymbol{\rm A}\boldsymbol{\rm \Lambda}\boldsymbol{\rm \Theta}\boldsymbol{\rm h}_{{\rm r}}\right|^2}{\sigma^2\|\boldsymbol{\rm h}^H\boldsymbol{\rm A}\boldsymbol{\rm \Lambda}\boldsymbol{\rm \Theta}\|^2+\delta^2})\geq R^{\min},~~{C_3}:\beta_{n}\in\{0,1\}.\\
\end{split}
\end{sequation}

With the given total power budget, define $P_{\rm I}=P_{\rm DC}+P_{\rm C}$, the optimization problem (\ref{eq62}) can be reduced to
\begin{sequation}\label{eq63}
\begin{split}
&\max\limits_{\mbox{\scriptsize$\begin{array}{c} 
		\boldsymbol{\rm \Theta},\rho,\beta_n
		\end{array}$}} 
\log_2(1+\frac{P^{\max}\left|\boldsymbol{\rm h}^H\boldsymbol{\rm A}\boldsymbol{\rm \Lambda}\boldsymbol{\rm \Theta}\boldsymbol{\rm h}_{{\rm r}}\right|^2}{\sigma^2\|\boldsymbol{\rm h}^H\boldsymbol{\rm A}\boldsymbol{\rm \Lambda}\boldsymbol{\rm \Theta}\|^2+\delta^2})\\
&s.t.~C_2,{C_3},{\bar C_1}: P^{\max}\|\boldsymbol{\rm A}\boldsymbol{\rm \Lambda}\boldsymbol{\rm \Theta}\boldsymbol{\rm h}_{{\rm r}}\|^2+\sigma^2\|\boldsymbol{\rm A}\boldsymbol{\rm \Lambda}\boldsymbol{\rm \Theta}\|_F^2\leq C-\sum_{n=1}^N\alpha_nP_{\rm I}.\\
\end{split}
\end{sequation}
\subsubsection{Phase-shift Optimization}Given the element on-off factor, it can be easily shown that the transmission rate in (\ref{eq63}) is maximized when the optimal phase-shift matrix of the RIS aligns the cascaded user-RIS-BS channel, i.e.,
\begin{sequation}\label{eq64}
\begin{split}
\theta_{n}=\arg([\boldsymbol{\rm h}]_n)-\arg([\boldsymbol{\rm h}_{\rm r}]_n),
\end{split}
\end{sequation}
where the optimal phase shift of the RIS has a similar form as the passive RIS. Then, by substituting the optimal phase shift in (\ref{eq64}) into the transmission rate expression and $\bar C_1$, the transmission rate expression and $\bar C_1$ are rewritten as
\begin{sequation}\label{eq65}
\begin{split}
&\frac{P^{\max}\left|\boldsymbol{\rm h}^H\boldsymbol{\rm A}\boldsymbol{\rm \Lambda}\boldsymbol{\rm \Theta}\boldsymbol{\rm h}_{{\rm r}}\right|^2}{\sigma^2\|\boldsymbol{\rm h}^H\boldsymbol{\rm A}\boldsymbol{\rm \Lambda}\boldsymbol{\rm \Theta}\|^2+\delta^2}=\frac{P^{\max}\left|\boldsymbol{\rm h}^H\boldsymbol{\rm A}\boldsymbol{\rm \Lambda}\boldsymbol{\rm \Theta}\boldsymbol{\rm h}_{{\rm r}}\right|^2}{\sigma^2|h|^2\sum\limits_{n=1}^N(\alpha_n\rho)^2+\delta^2}\overset{(a)}{=} \frac{P^{\max}|\sum\limits_{n=1}^N\alpha_n\rho|h||h_{\rm r}||^2}{\sigma^2|h|^2\sum\limits_{n=1}^N(\alpha_n\rho)^2+\delta^2}\overset{(b)}{\geq}\frac{P^{\max}|h|^2|h_{\rm r}|^2(\sum\limits_{n=1}^N\alpha_n\rho)^2}{\sigma^2|h|^2\sum\limits_{n=1}^N(\alpha_n\rho)^2+\delta^2},
\end{split}
\end{sequation}
\begin{sequation}\label{eq66}
\begin{split}
&P^{\max}\|\boldsymbol{\rm A}\boldsymbol{\rm \Lambda}\boldsymbol{\rm \Theta}\boldsymbol{\rm h}_{{\rm r}}\|^2+\sigma^2\|\boldsymbol{\rm A}\boldsymbol{\rm \Lambda}\boldsymbol{\rm \Theta}\|_F^2=P^{\max}|h_{\rm r}|^2\sum_{n=1}^N(\alpha_n\rho)^2+\sigma^2\sum_{n=1}^N(\alpha_n\rho)^2,
\end{split}
\end{sequation}
where $(a)$ utilizes the optimal design of $\boldsymbol{\rm \Theta}$, $(b)$ utilizes $|h|=\min\{|h_n|\}$ and $|h_{{\rm r}}|=\min\{|h_{{\rm r},n}|\}$. $h_n$ and $h_{{\rm r},n}$ are the $n$-th elements of $\boldsymbol{\rm h}$ and $\boldsymbol{\rm h}_{{\rm r}}$.
\subsubsection{Amplification Factor Optimization}
Next, we optimize the amplification factor. Define the number of reflecting elements in the active mode as $N_{\rm act}=\sum_{n=1}^N\alpha_n$, then the number of reflecting elements in the off mode is $N-N_{\rm act}$. Since
the logarithm function is a monotonic increasing function, the problem (\ref{eq63}) can be reduced to
\begin{sequation}\label{eq67}
\begin{split}
\max\limits_{\mbox{\scriptsize$\begin{array}{c} 
		\rho
		\end{array}$}} 
&\frac{P^{\max}|h|^2|h_{\rm r}|^2N_{\rm act}^2\rho^2}{\sigma^2|h|^2N_{\rm act}\rho^2+\delta^2}\\
s.t.~{\hat C_1}:& N_{\rm act}\rho^2(P^{\max}|h_{\rm r}|^2+\sigma^2)\leq C-N_{\rm act}P_{\rm I},~~\bar C_2:\frac{P^{\max}|h|^2|h_{\rm r}|^2N_{\rm act}^2\rho^2}{\sigma^2|h|^2N_{\rm act}\rho^2+\delta^2}\geq \bar R^{\min}.\\
\end{split}
\end{sequation}
In the following, we first drop ${\bar C_2}$ to make problem (\ref{eq67}) more treatable and then determine its feasible condition at the end. Since the SINR expression increases with $\rho$, the objective function can be maximized when $\rho$ takes its maximum value. Thus, $\hat C_1$ is active, the optimal $\rho$ is given by
\begin{sequation}\label{eq68}
\begin{split}
\rho=\sqrt{\frac{C-N_{\rm act}P_{\rm I}}{N_{\rm act}(P^{\max}|h_{\rm r}|^2+\sigma^2)}}.
\end{split}
\end{sequation}

\textbf{\textit{Remark 2:}} Based on (\ref{eq68}), as expected, the amplification factor increases with the increasing total power budget and decreases with the increasing number of reflecting elements in the active mode and the increasing power consumption of the active elements. This is because activating more reflecting elements or increasing the circuit power consumption of active elements will consume more power. 
\subsubsection{Number Configuration}
Next, we optimize the number of reflecting elements in the active mode. we substitute $\rho$ in (\ref{eq68}) into problem (\ref{eq67}), and problem (\ref{eq67}) can be reduced to
\begin{sequation}\label{eq69}
\begin{split}
\max\limits_{\mbox{\scriptsize$\begin{array}{c} 
		N_{\rm act}
		\end{array}$}} 
&\frac{P^{\max}|h|^2|h_{\rm r}|^2(CN_{\rm act}-P_{\rm I}N_{\rm act}^2)}{\sigma^2|h|^2(C-P_{\rm I}N_{\rm act})+\delta^2(P^{\max}|h_{\rm r}|^2+\sigma^2)}\\
s.t.~
\hat C_2:&\frac{P^{\max}|h|^2|h_{\rm r}|^2(CN_{\rm act}-P_{\rm I}N_{\rm act}^2)}{\sigma^2|h|^2(C-P_{\rm I}N_{\rm act})+\delta^2(P^{\max}|h_{\rm r}|^2+\sigma^2)}\geq \bar R^{\min}.\\
\end{split}
\end{sequation}
In the following, we also drop $\hat C_2$ to make problem (\ref{eq69}) more treatable and then determine its feasible condition at the end. Thus, we have
\begin{sequation}\label{eq70}
\begin{split}
\max\limits_{\mbox{\scriptsize$\begin{array}{c} 
		N_{\rm act}
		\end{array}$}} 
&\gamma=\frac{P^{\max}|h|^2|h_{\rm r}|^2(CN_{\rm act}-P_{\rm I}N_{\rm act}^2)}{\sigma^2|h|^2(C-P_{\rm I}N_{\rm act})+\delta^2(P^{\max}|h_{\rm r}|^2+\sigma^2)}.\\
\end{split}
\end{sequation}

\textbf{\textit{Proposition 1:}} The closed-form solution of $N_{\rm act}$ is given by
\begin{sequation}\label{eq71}
N_{\rm act}=\left\{
\begin{split}
&\min\{\left \lfloor N_{\rm act,1} \right \rfloor ,N\}, ~~0 \leq N< \left\lfloor N_{\rm act,2}\right\rfloor,\\
&{\rm arg} \max\limits_{\{N,\left \lfloor N_{\rm act,1} \right \rfloor \}} \gamma, ~~~~N\geq \left\lceil N_{\rm act,2}\right\rceil, 
\end{split}
\right.
\end{sequation}
where
\begin{sequation}\label{eq72}
\begin{split}
N_{\rm act,1}&=\frac{1}{\sigma^2|h|^2P_{\rm I}}\times \{(\sigma^2|h|^2C+\delta^2(P^{\max}|h_{\rm r}|^2+\sigma^2))\\
&-\sqrt{\delta^2(\sigma^2|h|^2C+\delta^2(P^{\max}|h_{\rm r}|^2+\sigma^2))(P^{\max}|h_{\rm r}|^2+\sigma^2)} \},
\end{split}
\end{sequation}
\begin{sequation}\label{eq73}
\begin{split}
N_{\rm act,2}&=\frac{1}{\sigma^2|h|^2P_{\rm I}}\times \{(\sigma^2|h|^2C+\delta^2(P^{\max}|h_{\rm r}|^2+\sigma^2))\\
&+\sqrt{\delta^2(\sigma^2|h|^2C+\delta^2(P^{\max}|h_{\rm r}|^2+\sigma^2))(P^{\max}|h_{\rm r}|^2+\sigma^2)} \}.
\end{split}
\end{sequation}
\begin{proof}
	Please refer to Appendix A.
\end{proof}
\subsubsection{Feasible Conditions for $N_{\rm act}$} So far, we have obtained the closed-form solutions for the phase shift, the amplification factor, and the number of active elements. In this subsection, we will discuss the feasible condition for $N_{\rm act}$. Define $a_1=P^{\max}|h|^2|h_{\rm r}|^2C+\sigma^2|h|^2P_{\rm I}\bar R^{\min}$ and $a_2=\sigma^2|h|^2C\bar R^{\min}+\delta^2\bar R^{\min}(P^{\max}|h_{\rm r}|^2+\sigma^2)$. Problem (\ref{eq67}) and  (\ref{eq69}) are feasible when the following conditions hold, i.e.,
\begin{sequation}\label{eq74}
\begin{split}
P^{\max}|h|^2|h_{\rm r}|^2C^2&\geq 4P_{\rm I}a_2,
\end{split}
\end{sequation}
and
\begin{sequation}\label{eq75}
\begin{split}
\max\{0,\left \lceil x_1 \right \rceil\}  \leq N_{\rm act}\leq   \left \lfloor x_2 \right \rfloor,
\end{split}
\end{sequation}
where
\begin{sequation}\label{eq76}
\begin{split}
&x_1=\frac{-a_1+\sqrt{a_1^2-4P^{\max}|h|^2|h_{\rm r}|^2P_{\rm I}a_2}}{-2P^{\max}|h|^2|h_{\rm r}|^2P_{\rm I}},
\end{split}
\end{sequation}
\begin{sequation}\label{eq77}
\begin{split}
&x_2=\frac{-a_1-\sqrt{a_1^2-4P^{\max}|h|^2|h_{\rm r}|^2P_{\rm I}a_2}}{-2P^{\max}|h|^2|h_{\rm r}|^2P_{\rm I}}.
\end{split}
\end{sequation}
\begin{proof}
	please refer to Appendix B.
\end{proof}

\subsection{Performance Comparison} 
In this subsection, we compare the performance of the two element on-off algorithms, i.e., how many passive elements for the passive RIS are needed to outperform the element on-off algorithm for the active RIS. Then, we have the following Proposition.

\textbf{\textit{Proposition 2:}}
Under the same total power budget, the element on-off algorithm for the passive RIS outperforms the element on-off algorithm for the active RIS if
\begin{sequation}\label{eq78}
\begin{split}
\frac{N_{\rm act}}{N_{\rm pas}^2}<\frac{\sigma^2|h|^2}{\delta^2}.
\end{split}
\end{sequation}
Otherwise, the element on-off algorithm for the active RIS performs better than that for the passive RIS when
\begin{sequation}\label{eq79}
\begin{split}
\frac{C-N_{\rm act}P_{\rm I}}{N_{\rm act}(P^{\max}|h_{\rm r}|^2+\sigma^2)}=\rho^2>\frac{1}{\frac{N_{\rm act}^2}{N_{\rm pas}^2}-\frac{\sigma^2|h|^2N_{\rm act}}{\delta^2}}.
\end{split}
\end{sequation}
\begin{proof}
	By using the solutions in (\ref{eq57}) and (\ref{eq65}), based on $\gamma_{\rm pas}>\gamma_{\rm act}$, we have
	 \begin{sequation}\label{eq80}
	 \begin{split}
	 &\frac{N_{\rm act}(P^{\max}|h_{\rm r}|^2+\sigma^2)}{C-N_{\rm act}P_{\rm I}}>\frac{N_{\rm act}^2}{N_{\rm pas}^2}-\frac{\sigma^2|h|^2N_{\rm act}}{\delta^2}\Rightarrow \frac{1}{\rho^2}>\frac{N_{\rm act}^2}{N_{\rm pas}^2}-\frac{\sigma^2|h|^2N_{\rm act}}{\delta^2}.
	 \end{split}
	 \end{sequation}
	 Since $\frac{1}{\rho^2}>0$, then the inequation in (\ref{eq80}) holds if the following inequation holds
	 \begin{sequation}\label{eq81}
	 \begin{split}
&\frac{N_{\rm act}^2}{N_{\rm pas}^2}-\frac{\sigma^2|h|^2N_{\rm act}}{\delta^2}<0\Rightarrow \frac{N_{\rm act}}{N_{\rm pas}^2}<\frac{\sigma^2|h|^2}{\delta^2},
	 \end{split}
	 \end{sequation}
which completes the proof.
\end{proof}
\textbf{\textit{Remark 3:}} From (\ref{eq81}), when the ratio of the number of active elements to  the square of  the number of passive elements is less than $\frac{\sigma^2|h|^2}{\delta^2}$, the element on-off algorithm for the passive RIS outperforms than that for the active RIS. This is because the number of passive elements is large enough and its performance is proportional to the number of passive elements, and the active RIS will suffer serious amplification noise. Furthermore, since the total power budget under both algorithms is the same, we have $N_{\rm act}<N_{\rm pas}$. Thus, (\ref{eq81}) still holds if $\frac{\sigma^2|h|^2N_{\rm pas}}{\delta^2}>1$, which means that a high amplification noise power, a low background noise power, or a large number of passive elements all contribute to better performance of the passive RIS.

\section{Simulation Results}
In this section, the effectiveness of the proposed algorithms is evaluated by comparing them with baseline algorithms. The proposed algorithms are defined as the proposed algorithm for the passive RIS and the proposed algorithm for the active RIS, respectively, and baseline algorithms are defined as
\begin{itemize}
	\item \textbf{The algorithms 1/2 for rate maximization}: These algorithms consider the objective function of maximizing the rate for the passive RIS and the active RIS, respectively.
	\item \textbf{The algorithms 1/2 with random phase shift}: These algorithms consider a random phase shift mechanism for the passive RIS and the active RIS, respectively.
	\item \textbf{Fully passive RIS}: All reflecting elements in a passive RIS are activated.
	\item \textbf{Fully active RIS}: All reflecting elements in an active RIS are activated.
\end{itemize}

A two-dimensional coordinate setup measured in meter (m) is considered, where the BS and the RIS are located at $(0, 0)$ m, $(50, 10)$ m, while the IoT devices are uniformly and randomly distributed in a circle centered at $(50, 0)$ m with a radius $3$ m.  The path-loss exponents for the BS-RIS links and the RIS-user links are $2.4$, $2.2$, respectively. The small-scale fading follows the Rayleigh distribution. Other parameters are: $P_k^{\max}\in[10,30]$ dBm, $R_{k}^{\min}=1$ bits/Hz, $P_{\rm RIS}^{\max}=10$ dBm, $\delta^2=-80$ dBm, $\sigma^2=-70$ dBm, $M=8$, $N=6$, $K=2$,   $P_{\rm C}=-10$ dBm, $P_{\rm DC}=-5$ dBm, $P_k=5$ dBm, $P^{\rm BS}=10$ dBm. 

\begin{figure}\vspace{-20pt}
	\centering
	\begin{subfigure}{0.35\linewidth}
		\centering
		\includegraphics[width=2in]{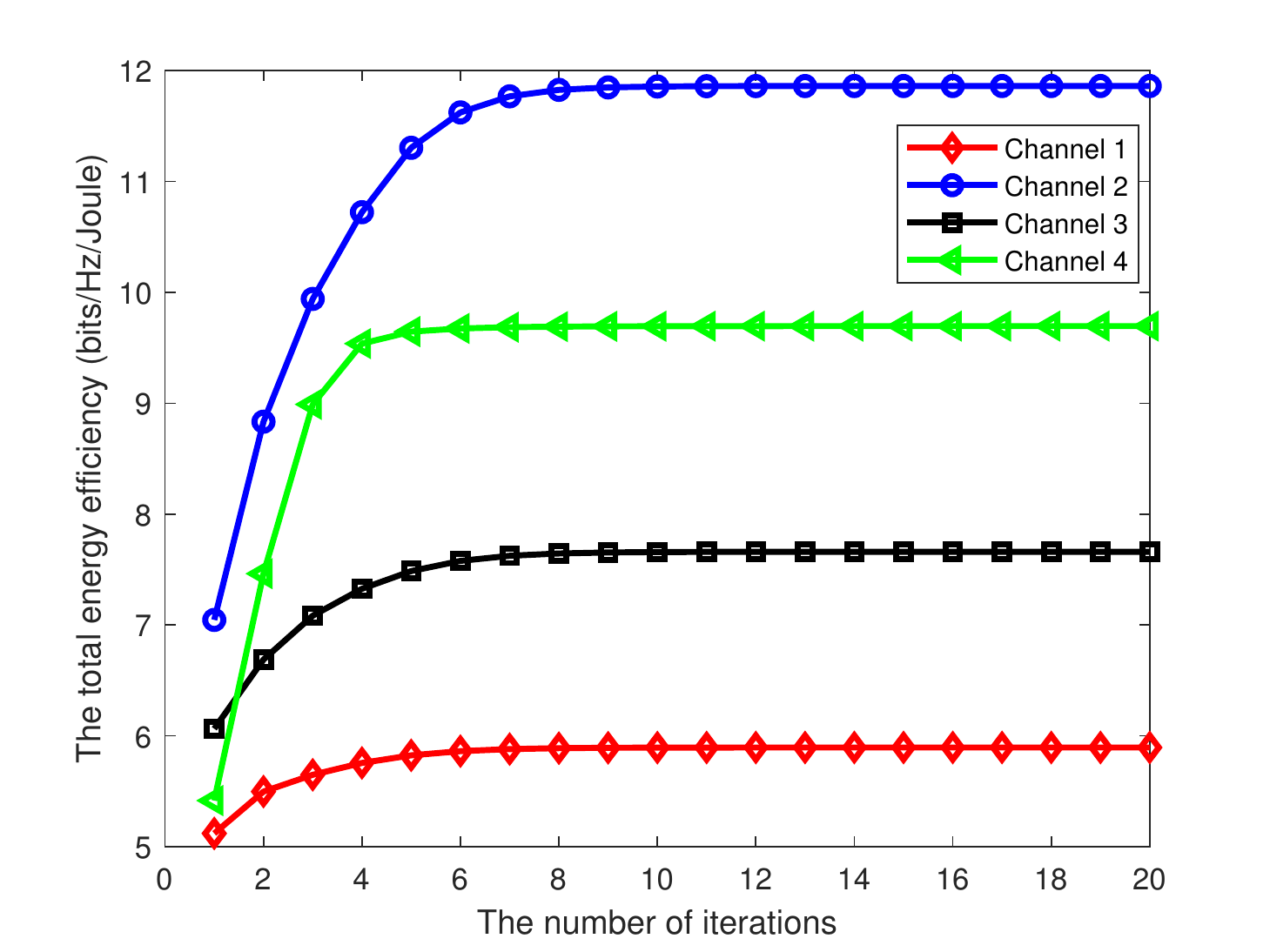}
		\caption{}
		\label{SRFig1}
	\end{subfigure}
	\centering
	\begin{subfigure}{0.35\linewidth}
		\centering
		\includegraphics[width=2in]{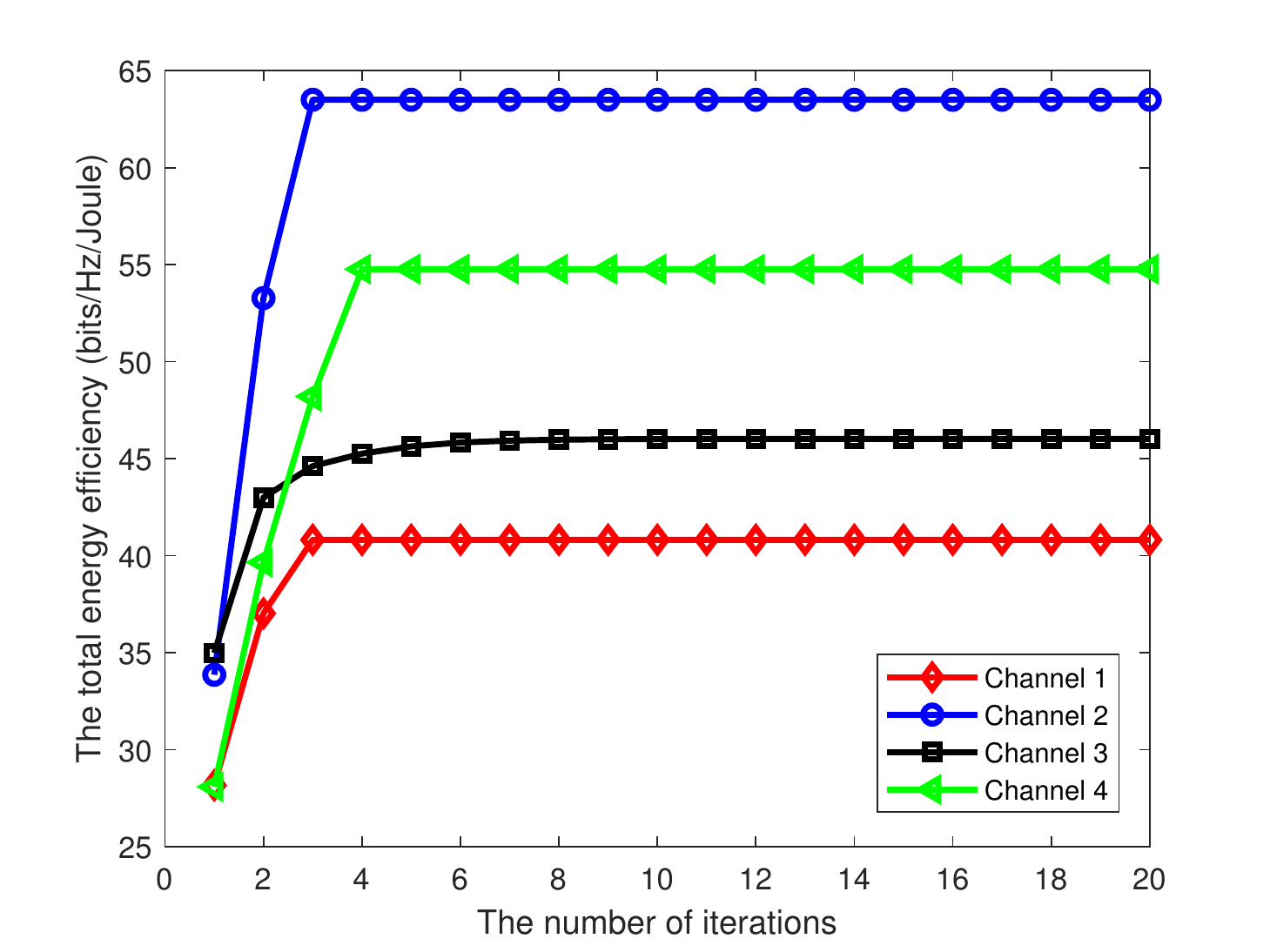}
		\caption{}
		\label{SRFig2}
	\end{subfigure}
	
	\begin{subfigure}{0.35\linewidth}
		\centering
		\includegraphics[width=2in]{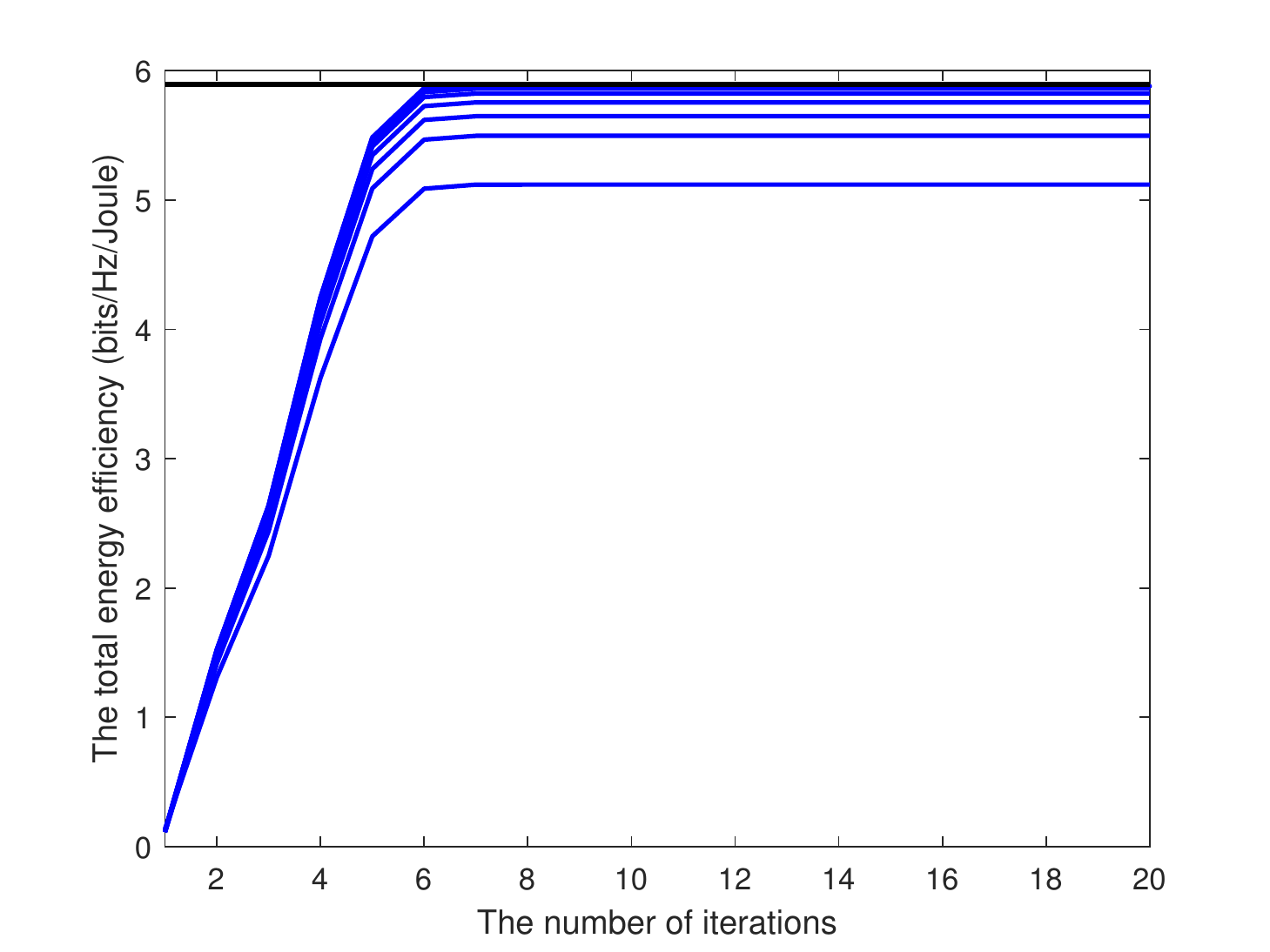}
		\caption{}
		\label{SRFig3}
	\end{subfigure}
	\centering
	\begin{subfigure}{0.35\linewidth}
		\centering
		\includegraphics[width=2in]{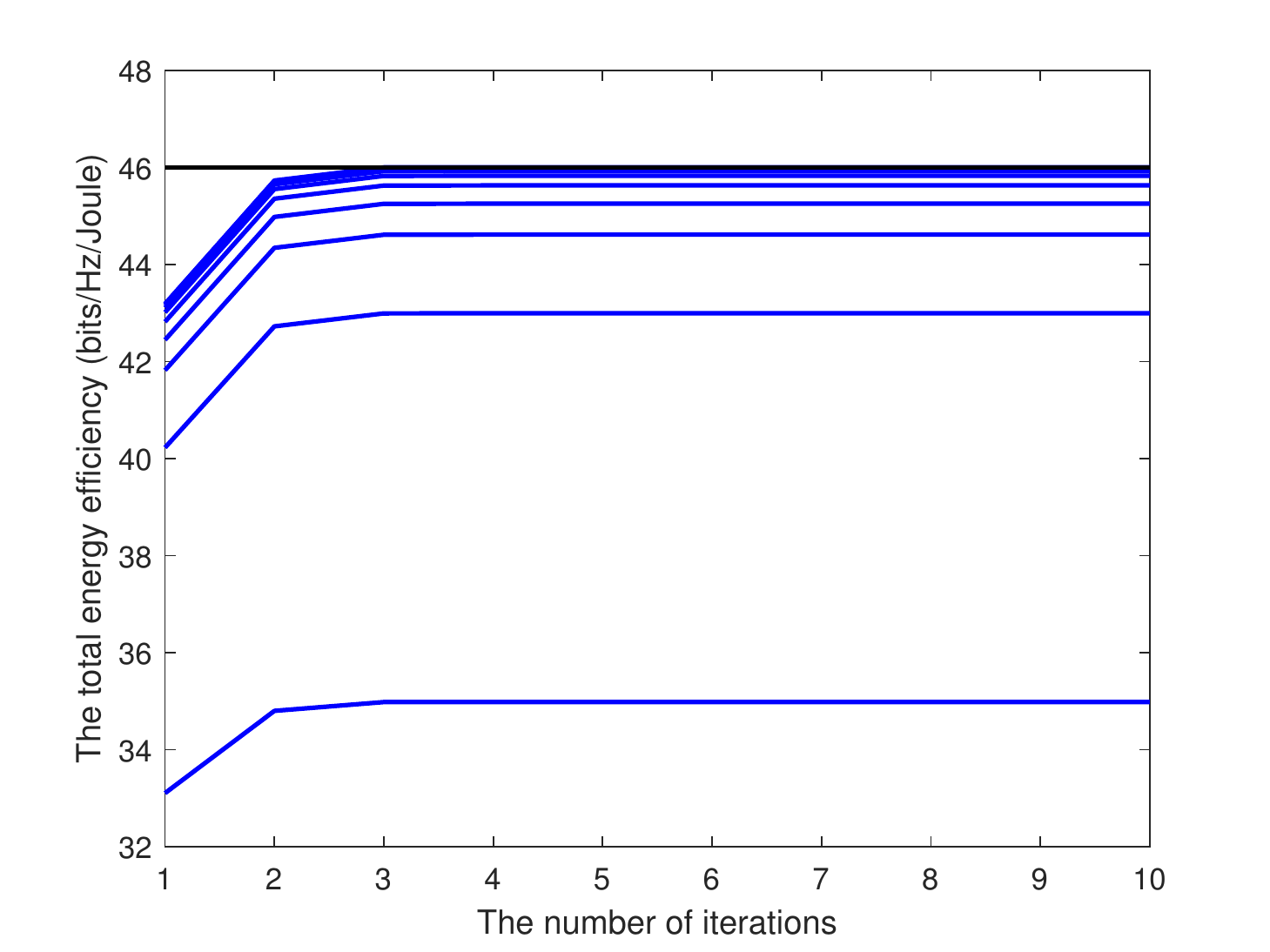}
		\caption{}
		\label{SRFig4}
	\end{subfigure}
	\caption{Convergence analysis for proposed algorithms. (a) Outer layer of the passive RIS. (b)  Outer layer of the active RIS. (c) Inner layer of the passive RIS. (d) Inner layer of the active RIS.}
	\label{SRFig1234}
\end{figure}

\subsection{Performance Analysis of Proposed Algorithms}
Fig. \ref{SRFig1234} evaluates the convergence performance of the proposed algorithms for the passive RIS and the active RIS. Specifically,  Fig. \ref{SRFig1234}(a) and Fig. \ref{SRFig1234}(b) give the convergence of the outer layer iteration under the proposed algorithms for the passive RIS and the active RIS under four arbitrarily selected channel realizations, respectively. It is observed that the proposed algorithms converge to the  stable value after several iterations, which reflects that the proposed algorithms have good convergence performance. Besides,  Fig. \ref{SRFig1234}(c) and Fig. \ref{SRFig1234}(d) give the convergence of the inner layer iteration under the proposed algorithms for passive RIS and active RIS, where the blue line represents the convergence status of each inner layer iteration, and the black line represents the stable value of solving the convex optimization problem (\ref{eq31}) and (\ref{eq53}) each time. It can be observed that after several iterations, each inner layer iteration converges to a stable value, and multiple inner layer iterations gradually approach the stable value of the outer layer iteration.

\begin{figure}[t]
	\vspace{-20pt}
	\centering
	\begin{minipage}[t]{0.48\textwidth}
		\centering
		\includegraphics[width=2.5in]{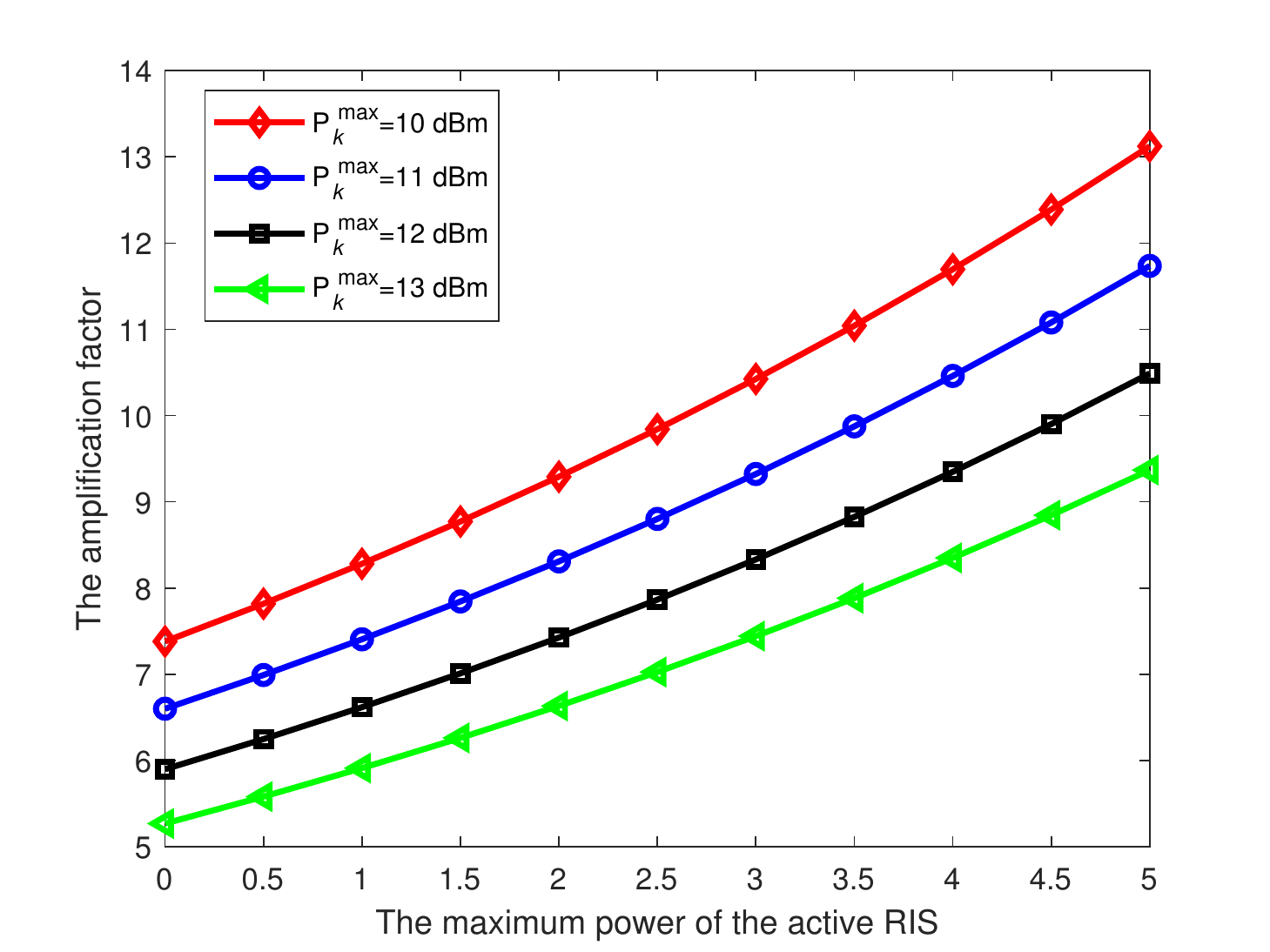}
		\caption{The amplification factor versus the maximum power of the acive RIS.}
		\label{SRFig5}
	\end{minipage}
	\begin{minipage}[t]{0.48\textwidth}
		\centering
		\includegraphics[width=2.5in]{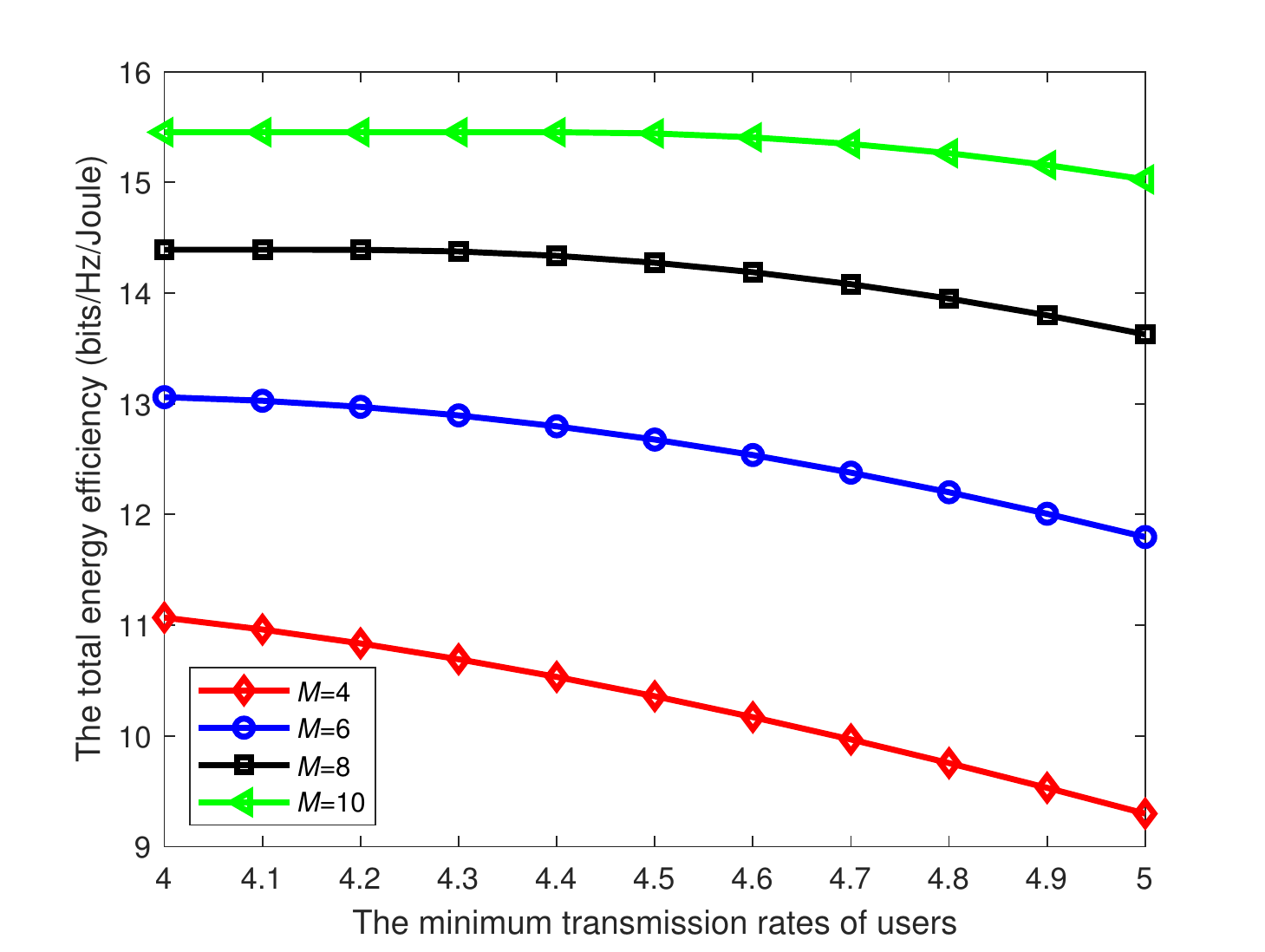}
		\caption{The total energy efficiency versus the minimum transmission rate of users.}
		\label{SRFig6}
	\end{minipage}
\end{figure}

Fig. \ref{SRFig5} shows the amplification factor versus the maximum power of the active RIS. Under the given reflecting element configuration, it can be observed that the amplification factor increases with the increasing maximum power of the active RIS. This is because the increase in the maximum power of the active RIS expands the feasible range of the amplification factor. Therefore, the system improves the energy efficiency by adopting a larger amplification factor. However, the amplification factor decreases with the increasing maximum transmit power of users since the increase in the maximum transmit power increases the transmit power of users and the energy efficiency gain brought by increasing the transmit power is higher than that of increasing the amplification factor. Therefore, the amplification factor can be decreased. This phenomenon can also be verified by the inverse proportionality between the amplification factor and the maximum transmit power in the formula (\ref{eq68}).

Fig. \ref{SRFig6} shows the total energy efficiency versus the minimum transmission rates of users under the proposed algorithm for the passive RIS. It can be observed that  the total energy efficiency increases with the increasing minimum transmission rate. Moreover, the decreasing trend is gentle when the minimum transmission rate is small, while it is sharp when the minimum transmission rate is large. This is because when the minimum transmission rate is small, the resource allocation strategy is in the optimal state. When the minimum transmission rate increases, the optimal state is broken. To satisfy the minimum transmission rate, the system will allocate higher transmit power and activate more reflecting elements, which, however, leads to a faster increase in total power consumption than in transmission rate, resulting in a decrease in total energy efficiency. Furthermore, the total energy efficiency increases with the increase in the number of BS antennas since the increase in the number of antennas provides additional spatial degrees of freedom. By coherently combining multi-path signals, it can increase the SINR and improve the total energy efficiency of the system.

\subsection{Performance Comparison of Proposed Algorithms}
\begin{figure}[t]
	\vspace{-20pt}
	\centering
	\begin{minipage}[t]{0.48\textwidth}
		\centering
		\includegraphics[width=2.5in]{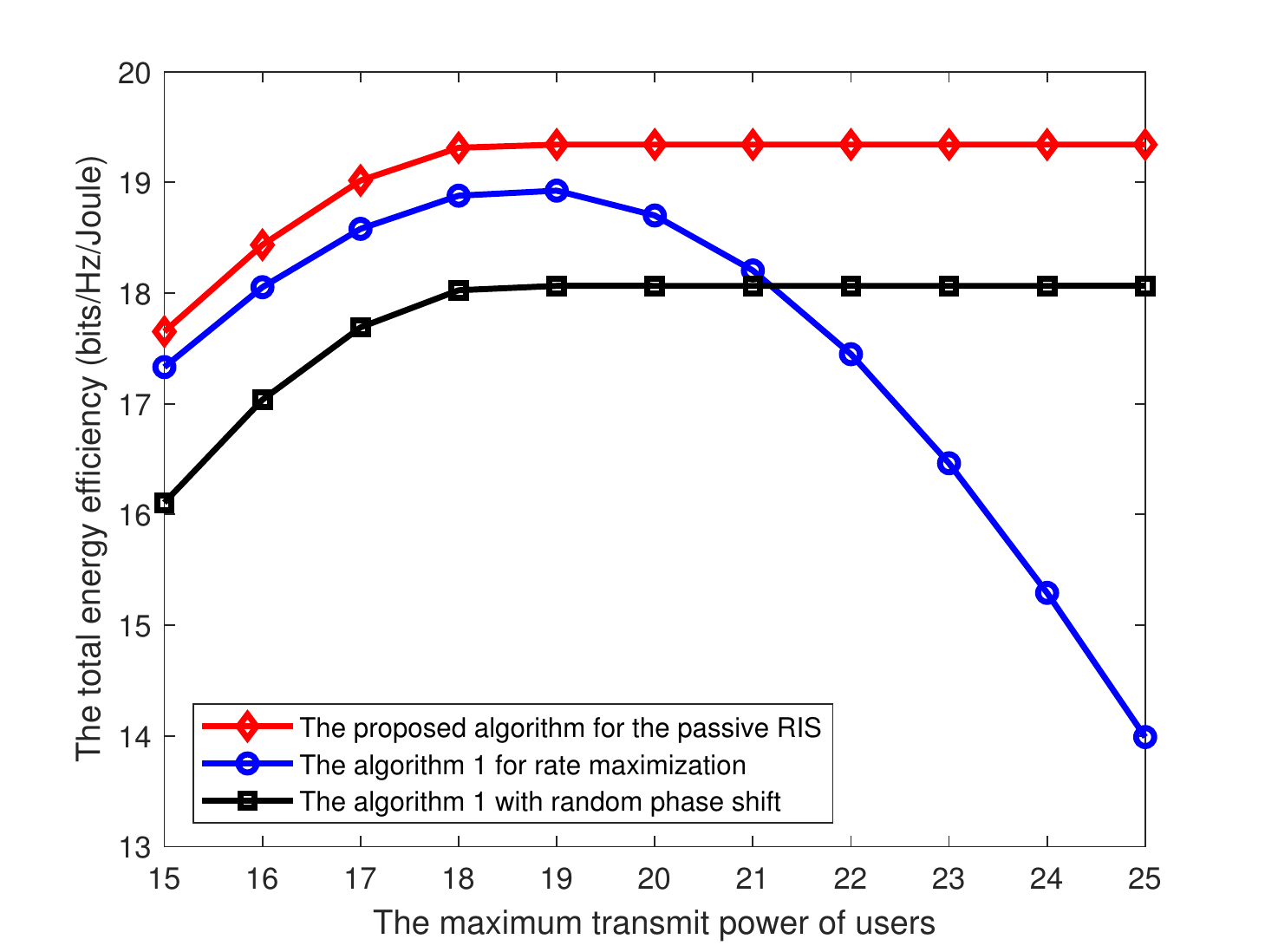}
		\caption{The total energy efficiency versus the maximum transmit power for the passive RIS.}
		\label{SRFig7-1}
	\end{minipage}
	\begin{minipage}[t]{0.48\textwidth}
		\centering
		\includegraphics[width=2.5in]{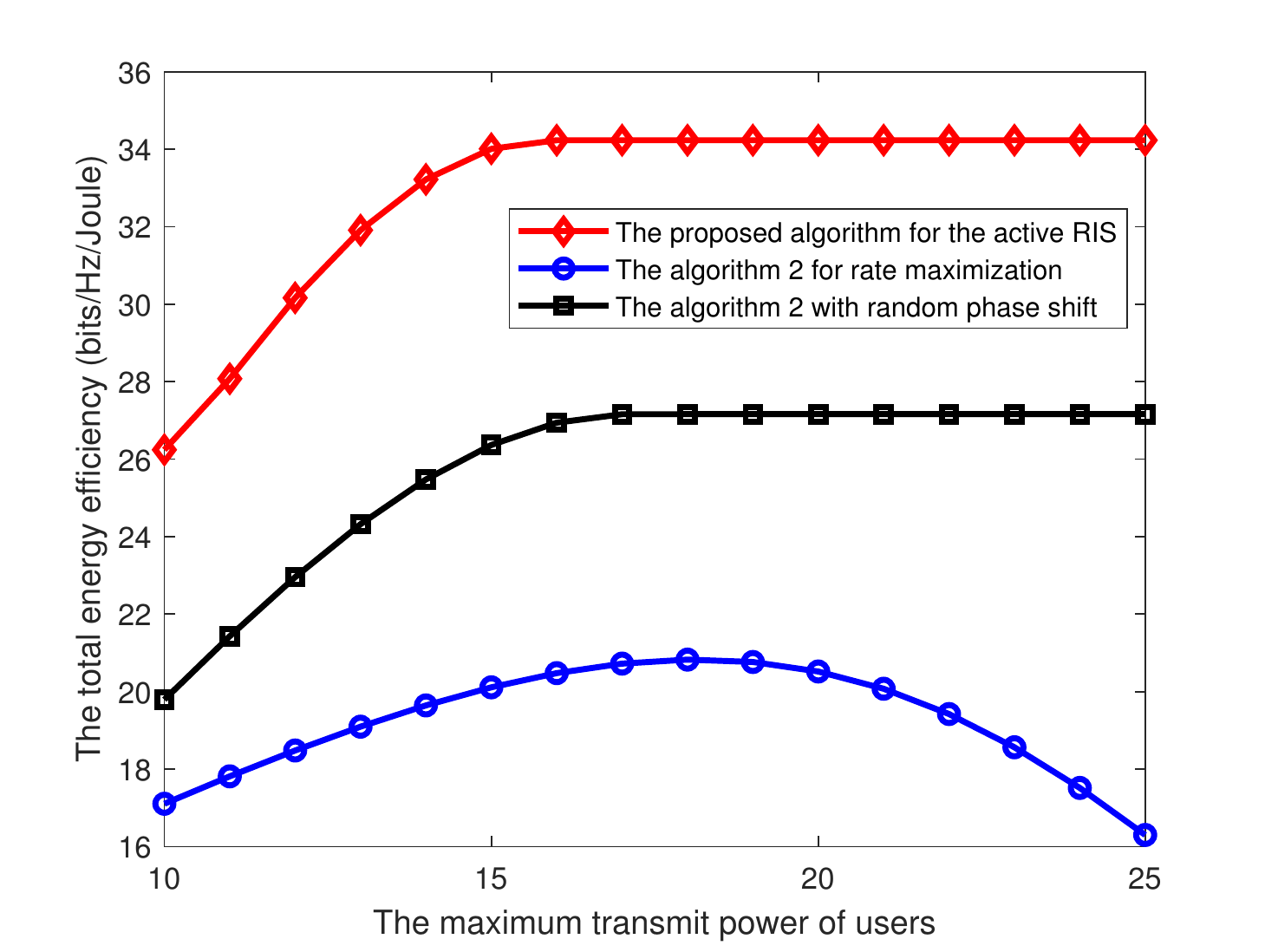}
		\caption{The total energy efficiency versus the maximum transmit power of users for the active RIS.}
		\label{SRFig7}
	\end{minipage}
\end{figure}

Fig. \ref{SRFig7-1} shows the total energy efficiency versus the maximum transmit power of users for the passive RIS under different algorithms. We can observe that the total energy efficiency under the proposed algorithm for the passive RIS and the algorithm 1 with random phase shift increases first and then gradually stabilizes with the increasing maximum transmit power of users. This is because when the maximum transmit power is small, the maximum transmit power constraint is active, and the maximum transmit power threshold is the optimal transmit power. The system continues to adjust the resource allocation strategy until the energy efficiency reaches the maximum value. Then, further increasing the transmit power will reduce the total energy efficiency. Therefore, the system will no longer adjust the resource allocation strategy. However, the total energy efficiency of the algorithm 1 for rate maximization increases first and then decreases with the increasing maximum transmit power since the algorithm 1 for rate maximization only focuses on improving the transmission rate, neglecting the total power consumption. From the expression of the transmission rate, it can be seen that the optimal transmit power is always equal to the maximum transmit power threshold. When the energy efficiency reaches the optimal value, increasing the transmit power further increases the total power consumption of the system, thereby reducing the total energy efficiency. 

Fig. \ref{SRFig7} shows the total energy efficiency versus the maximum transmit power of users for the active RIS under different algorithms.  We can observe a phenomenon similar to that in Fig. \ref{SRFig7-1}, and the reason for this is similar for both the active RIS and the passive RIS. It is worth noting that when the maximum transmit power is small, the total energy efficiency of the algorithm 2 for rate maximization is the worst. Unlike in the passive RIS, the rate maximization algorithm in active RIS will activate all reflecting elements and set the amplification factor to the maximum value or reach the maximum power threshold of the active RIS, which will result in high power consumption and low total energy efficiency.

\begin{figure}\vspace{-20pt}
	\centering
	\includegraphics[width=2.5in]{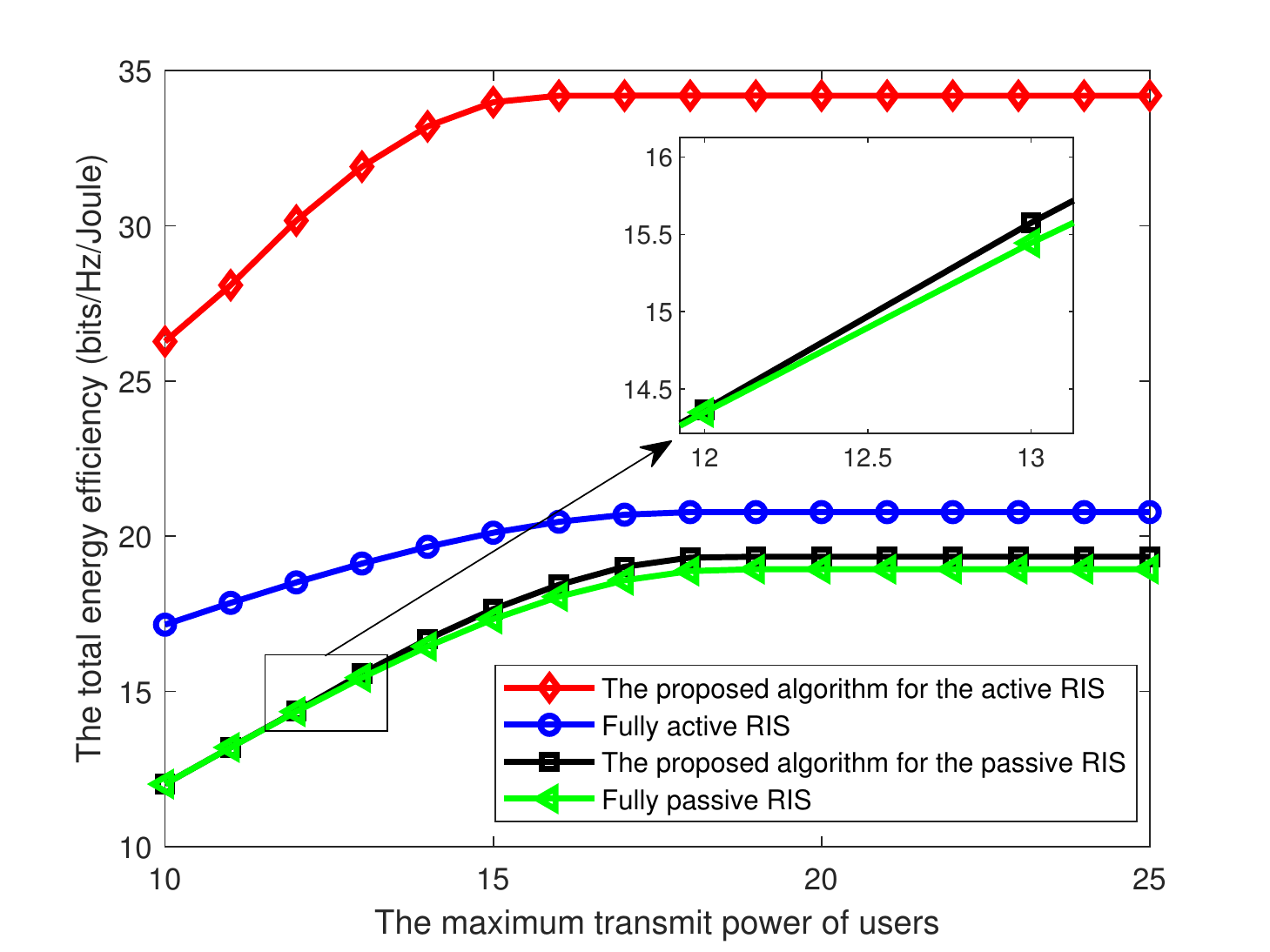}
	\caption{The total energy efficiency versus the maximum transmit power for the proposed algorithms.}
	\label{SRFig8}
\end{figure}

Fig. \ref{SRFig8} shows the total energy efficiency versus the maximum transmit power of users under different algorithms.  From Fig. \ref{SRFig8}, it can be seen that the proposed algorithm for the active RIS outperforms other algorithms in terms of energy efficiency. On the one hand, the energy efficiency gain brought by active RIS is greater than the energy efficiency loss caused by its power consumption. On the other hand, the fully active algorithm activates all reflecting elements, which, however, leads to an unnecessary increase in power consumption and a decrease in energy efficiency. The proposed algorithms can flexibly adjust the activation and deactivation of the reflecting elements, thus achieving maximum energy efficiency. Besides, when the maximum transmit power of users is small, the proposed algorithm for the passive RIS has the same energy efficiency as the fully passive RIS algorithm. However, with the increasing maximum transmit power, the proposed algorithm for the passive RIS outperforms the fully passive algorithm. This is because when the maximum transmit power is small, the proposed algorithm for the passive RIS activates all reflecting elements to meet the user's QoS. However, when the maximum transmit power is large, the proposed algorithm for the passive RIS closes some elements to improve the total energy efficiency, which reflects the flexibility of the proposed algorithms.

\begin{figure*}[t]\vspace{-20pt}
	\centering
	\begin{subfigure}{0.49\linewidth}
		\centering
		\includegraphics[width=2.5in]{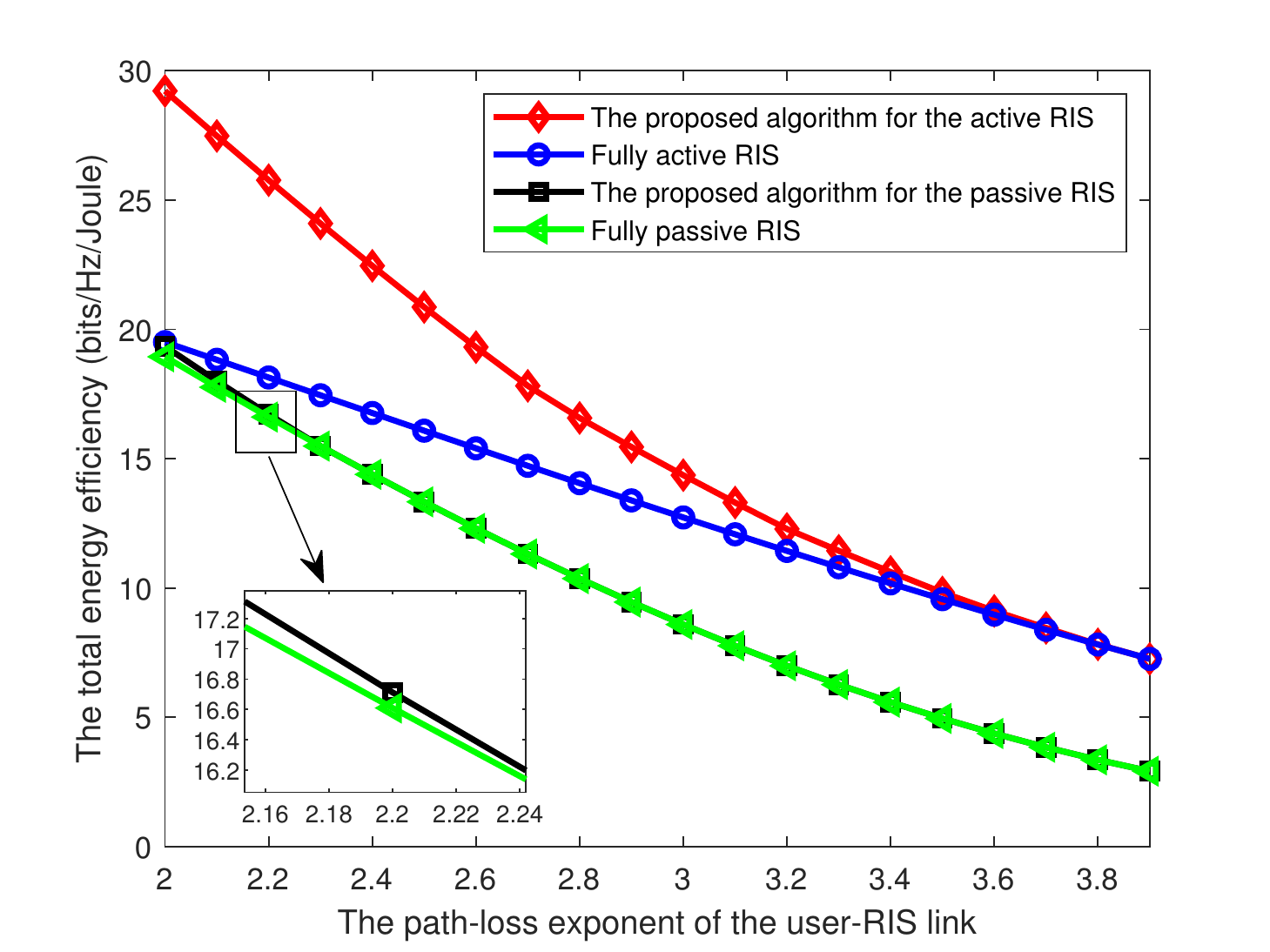}
		\caption{The total energy efficiency versus the path-loss exponent.}
		\label{SRFig9a}
	\end{subfigure}
	\centering
	\begin{subfigure}{0.49\linewidth}
		\centering
		\includegraphics[width=2.5in]{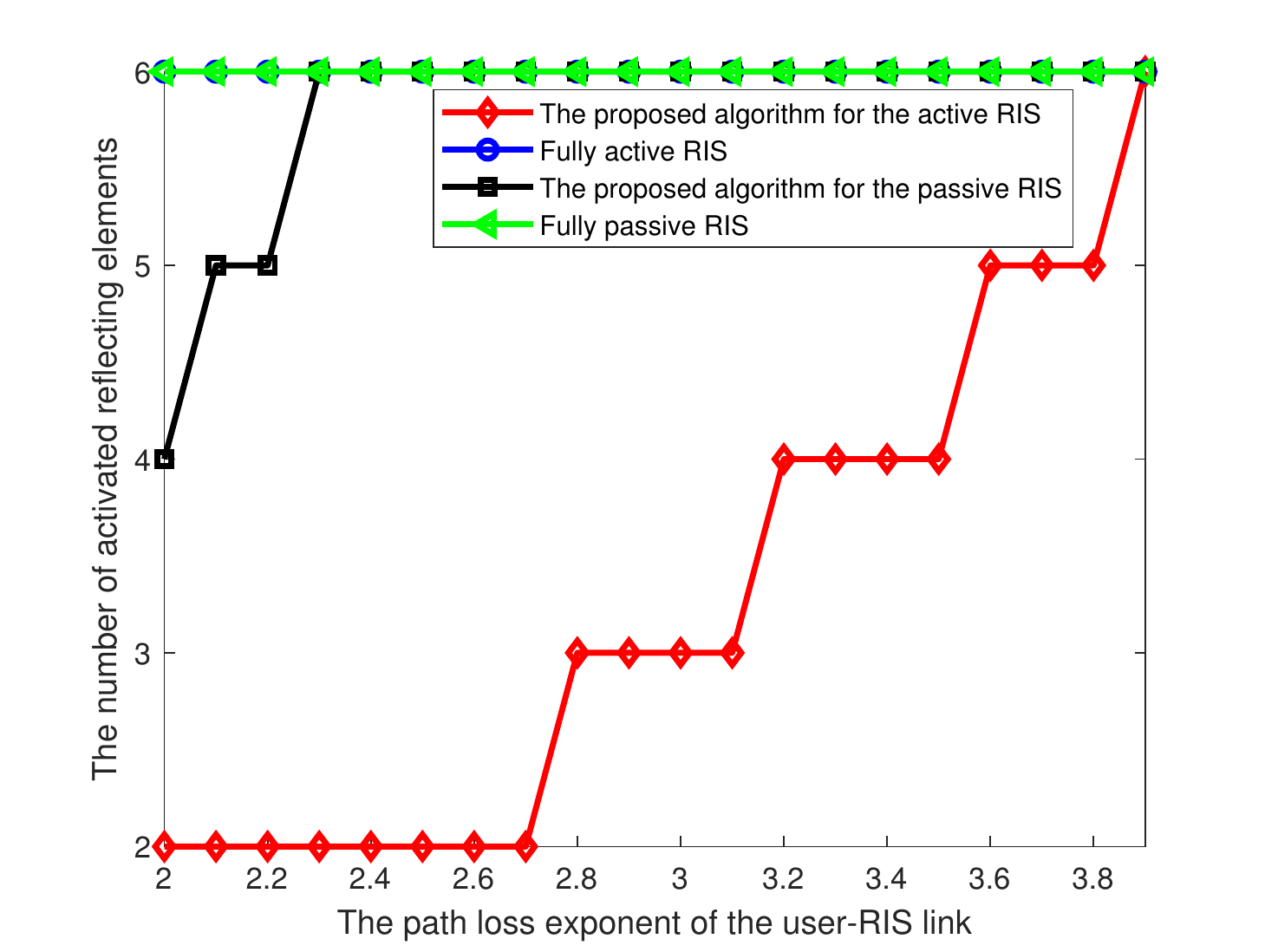}
		\caption{The number of activated reflecting elements versus the path-loss exponent.}
		\label{SRFig9b}
	\end{subfigure}
	\caption{The total energy efficiency and the number of activated reflecting elements versus the path-loss exponent.}
	\label{SRFig9}
\end{figure*}

Fig. \ref{SRFig9} shows the total energy efficiency and the number of activated reflecting elements versus the path-loss exponent under different algorithms.  From Fig. \ref{SRFig9}(a), it can be seen that with the increasing path-loss exponent, the total energy efficiency under all algorithms decreases. This is because the larger the path-loss exponent, the worse the channel environment, which undoubtedly reduces the total energy efficiency. In addition, when the path-loss exponent is small, the proposed algorithm for the active RIS outperforms the fully active RIS algorithm, while when the path-loss exponent is large, the proposed algorithm for the active RIS has the same performance as the fully active RIS algorithm. This is because when the path-loss exponent is small, the channel conditions are good, and activating all reflecting elements will only increase power consumption, thereby reducing the total energy efficiency. Therefore, the proposed algorithm for the active RIS will not activate all reflecting elements to improve energy efficiency, as shown in Fig. \ref{SRFig9}(b). However, when the path-loss exponent is large, the proposed algorithm for the active RIS needs to activate all reflecting elements to meet the user's QoS. The proposed algorithm for the passive RIS exhibits similar phenomena to the active RIS, as shown in Fig. \ref{SRFig9}(a) and Fig. \ref{SRFig9}(b). Overall, the proposed algorithms always outperform existing algorithms or have the same performance as existing algorithms, which reflects the flexibility and effectiveness of the proposed algorithms.

\section{Conclusion}
This paper proposes a novel element on-off mechanism and analyzes the impact of the element on-off mechanism on passive RIS and active RIS-aided wireless networks. In particular, two different problems for the passive RIS and the active RIS are respectively formulated with the objective of maximizing total energy efficiency while satisfying constraints on the maximum transmit power of users and the active RIS, the QoS requirement, and the phase-shift matrix. Due to the intractable problems, we respectively develop two AO-based iterative algorithms applying quadratic transform, Big-M formulation, and the SCA method to deal with the highly non-convex problems.  Moreover, we reduce the original problem to rate maximization problems for giving the same total power budget, explore the tradeoff between the number of activating and deactivating elements, and compare the performance of the proposed algorithms for the passive RIS and the active RIS. Simulation results demonstrate that the proposed algorithms outperform baseline algorithms and can flexibly activate reflecting elements.

\appendices

\section{}
we substitute $\rho$ in (\ref{eq68}) into the SINR expression, i.e.,
\begin{sequation}\label{eq82}
\begin{split}
\gamma=\frac{P^{\max}|h|^2|h_{\rm r}|^2(CN_{\rm act}-P_{\rm I}N_{\rm act}^2)}{\sigma^2|h|^2(C-P_{\rm I}N_{\rm act})+\delta^2(P^{\max}|h_{\rm r}|^2+\sigma^2)}.
\end{split}
\end{sequation}
The first-order derivative of $\gamma$ in (\ref{eq82}) with respect to
$N_{\rm act}$ is given by
\begin{sequation}\label{eq83}
\begin{split}
\frac{\partial \gamma}{\partial N_{\rm act}}=\frac{P^{\max}|h|^2|h_{\rm r}|^2\times f_1(N_{\rm act})}{(\sigma^2|h|^2(C-P_{\rm I}N_{\rm act})+\delta^2(P^{\max}|h_{\rm r}|^2+\sigma^2))^2},
\end{split}
\end{sequation}
where
\begin{sequation}\label{eq84}
\begin{split}
f_1(N_{\rm act})&=\sigma^2|h|^2P_{\rm I}^2N_{\rm act}^2-2P_{\rm I}(\sigma^2|h|^2C+\delta^2(P^{\max}|h_{\rm r}|^2+\sigma^2))N_{\rm act}+C(\sigma^2|h|^2C+\delta^2(P^{\max}|h_{\rm r}|^2+\sigma^2)).
\end{split}
\end{sequation}
Then, we can find
\begin{sequation}\label{eq85}
\begin{split}
&4P_{\rm I}^2(\sigma^2|h|^2C+\delta^2(P^{\max}|h_{\rm r}|^2+\sigma^2))^2-4\sigma^2|h|^2P_{\rm I}^2C(\sigma^2|h|^2C+\delta^2(P^{\max}|h_{\rm r}|^2+\sigma^2))\\
=&4P_{\rm I}^2\delta^2(\sigma^2|h|^2C+\delta^2(P^{\max}|h_{\rm r}|^2+\sigma^2))(P^{\max}|h_{\rm r}|^2+\sigma^2)> 0.
\end{split}
\end{sequation}
Therefore, there must exist two roots, i.e.,
\begin{sequation}\label{eq86}
\begin{split}
&N_{\rm act,1}=\frac{1}{\sigma^2|h|^2P_{\rm I}}\times \{(\sigma^2|h|^2C+\delta^2(P^{\max}|h_{\rm r}|^2+\sigma^2))-\\
&\sqrt{\delta^2(\sigma^2|h|^2C+\delta^2(P^{\max}|h_{\rm r}|^2+\sigma^2))(P^{\max}|h_{\rm r}|^2+\sigma^2)} \},
\end{split}
\end{sequation}
\begin{sequation}\label{eq87}
\begin{split}
&N_{\rm act,2}=\frac{1}{\sigma^2|h|^2P_{\rm I}}\times \{(\sigma^2|h|^2C+\delta^2(P^{\max}|h_{\rm r}|^2+\sigma^2))+\\
&\sqrt{\delta^2(\sigma^2|h|^2C+\delta^2(P^{\max}|h_{\rm r}|^2+\sigma^2))(P^{\max}|h_{\rm r}|^2+\sigma^2)} \}.
\end{split}
\end{sequation}
Furthermore, $P_{\rm I}(\sigma^2|h|^2C+\delta^2(P^{\max}|h_{\rm r}|^2+\sigma^2))=\sqrt{P_{\rm I}^2(\sigma^2|h|^2C+\delta^2(P^{\max}|h_{\rm r}|^2+\sigma^2))^2}\geq\sqrt{\delta^2P_{\rm I}^2(\sigma^2|h|^2C+\delta^2(P^{\max}|h_{\rm r}|^2+\sigma^2))(P^{\max}|h_{\rm r}|^2+\sigma^2)}$. Thus, both two roots are positive, and $0<N_{\rm act,1}<N_{\rm act,2}$. We can obtained that $\gamma$ increases when $0\leq N_{\rm act} <\left\lfloor N_{\rm act,1}\right\rfloor$ and $N_{\rm act} > \left\lceil N_{\rm act,2}\right\rceil$, and $\gamma$ decreases when $\left\lceil N_{\rm act,1}\right\rceil\leq N_{\rm act} <\left\lfloor  N_{\rm act,2}\right\rfloor$. Then, we have the following cases: $N_{\rm act}=\min\{\left \lfloor N_{\rm act,1} \right \rfloor ,N\}$ when $0 \leq N< \left\lfloor N_{\rm act,2}\right\rfloor$; $N_{\rm act}={\rm arg} \max\limits_{\{N,\left \lfloor N_{\rm act,1} \right \rfloor \}} \gamma$ when $ N\geq \left\lceil N_{\rm act,2}\right\rceil$. 

The proof is completed.

\section{}
For optimal $N_{\rm act}$, the minimum transmission rate constraint must be satisfied. Thus, we substitute $\rho$ into the minimum transmission rate constraint, i.e., 
\begin{sequation}\label{eq88}
\begin{split}
\frac{P^{\max}|h|^2|h_{\rm r}|^2(CN_{\rm act}-P_{\rm I}N_{\rm act}^2)}{\sigma^2|h|^2(C-P_{\rm I}N_{\rm act})+\delta^2(P^{\max}|h_{\rm r}|^2+\sigma^2)}\geq \bar R^{\min}.
\end{split}
\end{sequation}
After some simplifications, we have 
\begin{sequation}\label{eq89}
\begin{split}
&-P^{\max}|h|^2|h_{\rm r}|^2P_{\rm I}N_{\rm act}^2+a_1N_{\rm act}-a_2\geq 0,
\end{split}
\end{sequation}
where $a_1=P^{\max}|h|^2|h_{\rm r}|^2C+\sigma^2|h|^2P_{\rm I}\bar R^{\min}$ and $a_2=\sigma^2|h|^2C\bar R^{\min}+\delta^2\bar R^{\min}(P^{\max}|h_{\rm r}|^2+\sigma^2)$.
Then, the minimum transmission rate can be guaranteed when the following conditions hold
\begin{sequation}\label{eq90}
\begin{split}
&a_1^2-4P^{\max}|h|^2|h_{\rm r}|^2P_{\rm I}a_2\geq 0\Rightarrow (P^{\max}|h|^2|h_{\rm r}|^2C)^2-4P^{\max}|h|^2|h_{\rm r}|^2P_{\rm I}a_2\geq 0\Rightarrow P^{\max}|h|^2|h_{\rm r}|^2C^2\geq 4P_{\rm I}a_2.
\end{split}
\end{sequation}
Then, there exist two roots, i.e.,
\begin{sequation}\label{eq91}
\begin{split}
&x_1=\frac{-a_1+\sqrt{a_1^2-4P^{\max}|h|^2|h_{\rm r}|^2P_{\rm I}a_2}}{-2P^{\max}|h|^2|h_{\rm r}|^2P_{\rm I}},~~
x_2=\frac{-a_1-\sqrt{a_1^2-4P^{\max}|h|^2|h_{\rm r}|^2P_{\rm I}a_2}}{-2P^{\max}|h|^2|h_{\rm r}|^2P_{\rm I}}.
\end{split}
\end{sequation}
Thus, the minimum transmission rate can be guaranteed if optimal $N_{\rm act}$ satisfies the following condition
\begin{sequation}\label{eq93}
\begin{split}
\max\{0,\left \lceil x_1 \right \rceil\}  \leq N_{\rm act}\leq   \left \lfloor x_2 \right \rfloor.
\end{split}
\end{sequation}
The proof is completed.


\begin{thebibliography}{99}
	
	\bibitem{a1}
	Y. Liu \textit{et al.}, ``Rethinking sustainable sensing in agricultural Internet of Things: From power supply perspective,'' \textit{IEEE Wireless Commun.}, vol. 29, no. 4, pp. 102--109, Aug. 2022.
	
	\bibitem{a2}
	Y. Liu \textit{et al.}, ``Understanding the impact of environmental conditions on zero-power Internet of Things: An experimental evaluation,'' \textit{IEEE Wireless Commun.}, doi: 10.1109/MWC.007.2200177.
	
	\bibitem{b1}
	C. Pan \textit{et al.}, ``Reconfigurable intelligent surfaces for 6G systems: Principles, applications, and research directions,'' \textit{IEEE Commun. Mag.}, vol. 59, no. 6, pp. 14--20, Jun. 2021.
	
	\bibitem{b2}
	D. Li, ``Ergodic capacity of intelligent reflecting surface-assisted communication systems with phase errors,''  \textit{IEEE Commun. Lett.}, vol. 24, no. 8, pp. 1646--1650, Aug. 2020.
	
	\bibitem{c1}
	Y. Yang, S. Zhang, and R. Zhang, ``IRS-enhanced OFDMA: Joint resource allocation and passive beamforming optimization,'' \textit{IEEE Wireless Commun. Lett.}, vol. 9, no. 6, pp. 760--764, Jun. 2020.
	
	\bibitem{c2}
	H. Xie \textit{et al.}, ``Gain without pain: Recycling reflected energy from wireless powered RIS-aided communications,'' \textit{IEEE Internet Things J.}, doi: 10.1109/JIOT.2023.3262517.
	
	\bibitem{c3}
	Q. -U. -A. Nadeem \textit{et al.}, ``Intelligent reflecting surface-assisted multi-user MISO communication: Channel estimation and beamforming design,'' \textit{IEEE Open J. Commun. Soc.}, vol. 1, pp. 661--680, 2020.
	
	\bibitem{c4}
	M. Fu and R. Zhang, ``Active and passive IRS jointly aided communication: Deployment design and achievable rate,'' \textit{IEEE Wireless Commun. Lett.}, vol. 12, no. 2, pp. 302--306, Feb. 2023.
	
	\bibitem{c5}
	X. Mu, Y. Liu, L. Guo, J. Lin, and R. Schober, ``Joint deployment and multiple access design for intelligent reflecting surface assisted networks,'' \textit{IEEE Trans. Wireless Commun.}, vol. 20, no. 10, pp. 6648--6664, Oct. 2021.
	
	\bibitem{c5-1}
	Z. Kang, C. You, and R. Zhang, ``IRS-aided wireless relaying: Deployment strategy and capacity scaling,'' \textit{IEEE Wireless Commun. Lett.}, vol. 11, no. 2, pp. 215--219, Feb. 2022.
	
	\bibitem{c6}
	B. Zheng, C. You, and R. Zhang, ``Double-IRS assisted multi-user MIMO: Cooperative passive beamforming design,'' \textit{IEEE Trans. Wireless Commun.}, vol. 20, no. 7, pp. 4513--4526, Jul. 2021.
	
	\bibitem{c7}
	Y. Xu, H. Xie, Q. Wu, C. Huang, and C. Yuen, ``Robust max-min energy efficiency for RIS-aided HetNets with distortion noises,'' \textit{IEEE Trans. Commun.}, vol. 70, no. 2, pp. 1457--1471, Feb. 2022.
	
	\bibitem{c8}
	T. V. Nguyen, D. N. Nguyen, M. D. Renzo, and R. Zhang, ``Leveraging secondary reflections and mitigating interference in multi-IRS/RIS aided wireless networks,'' \textit{IEEE Trans. Wireless Commun.}, vol. 22, no. 1, pp. 502--517, Jan. 2023.
	
	\bibitem{c9}
	W. Mei and R. Zhang, ``Cooperative beam routing for multi-IRS aided communication,'' \textit{IEEE Wireless Commun. Lett.}, vol. 10, no. 2, pp. 426--430, Feb. 2021.

	
	\bibitem{d1}
	R. Long, Y. -C. Liang, Y. Pei, and E. G. Larsson, ``Active reconfigurable intelligent surface-aided wireless communications,'' \textit{IEEE Trans. Wireless Commun.}, vol. 20, no. 8, pp. 4962--4975, Aug. 2021.
	
	\bibitem{d2}
	C. -W. Chen, W. -C. Tsai, and A. -Y. Wu, ``Low-complexity two-step optimization in active-IRS-assisted uplink NOMA communication,'' \textit{IEEE Commun. Lett.}, vol. 26, no. 12, pp. 2989--2993, Dec. 2022.
	
	\bibitem{d3}
	P. Zeng, D. Qiao, Q. Wu, and Y. Wu, ``Throughput maximization for active intelligent reflecting surface-aided wireless powered communications,'' \textit{IEEE Wireless Commun. Lett.}, vol. 11, no. 5, pp. 992-996, May 2022.
	
	\bibitem{d4}
	L. Dong, H. -M. Wang, and J. Bai, ``Active reconfigurable intelligent surface aided secure transmission,'' \textit{IEEE Trans. Veh. Technol.}, vol. 71, no. 2, pp. 2181--2186, Feb. 2022.
	
	\bibitem{d5}
	H. Niu \textit{et al.}, ``Active RIS-assisted secure transmission for cognitive satellite terrestrial networks,'' \textit{IEEE Trans. Veh. Technol.}, vol. 72, no. 2, pp. 2609--2614, Feb. 2023.
	
	\bibitem{d6}
	Y. Gao, Q. Wu, G. Zhang, W. Chen, D. W. K. Ng, and M. D. Renzo, ``Beamforming optimization for active intelligent reflecting surface-aided SWIPT,'' \textit{IEEE Trans. Wireless Commun.}, vol. 22, no. 1, pp. 362--378, Jan. 2023.
	
	
	\bibitem{d7}
	Z. Zhang \textit{et al.}, ``Active RIS vs. Passive RIS: Which will prevail in 6G?,'' \textit{IEEE Trans. Commun.}, vol. 71, no. 3, pp. 1707--1725, Mar. 2023.
	
	\bibitem{d8}
	R. Su, L. Dai, J. Tan, M. Hao, and R. MacKenzie, ``Capacity enhancement for reconfigurable intelligent surface-aided wireless network: From regular array to irregular array,'' \textit{IEEE Trans. Veh. Technol.}, doi: 10.1109/TVT.2023.3236179.
	
	\bibitem{d9}
	K. Zhi, C. Pan, H. Ren, K. K. Chai, and M. Elkashlan, ``Active RIS versus passive RIS: Which is superior with the same power budget?,'' \textit{IEEE Commun. Lett.}, vol. 26, no. 5, pp. 1150--1154, May 2022.
	
	\bibitem{d10}
	D. Li, ``How many reflecting elements are needed for energy- and spectral-efficient intelligent reflecting surface-assisted communication,'' \textit{IEEE Trans. Commun.}, vol. 70, no. 2, pp. 1320--1331, Feb. 2022.
	
	\bibitem{e1}
	J. -C. Chen, ``Capacity improvement for intelligent reflecting surface-assisted wireless systems with a limited number of passive elements,'' \textit{IEEE Wireless Commun. Lett.}, vol. 11, no. 4, pp. 801-805, Apr. 2022.
	
	\bibitem{h1}
	D. Li, ``Fairness-aware multiuser scheduling for finite-resolution intelligent reflecting surface-assisted communication,'' \textit{IEEE Commun. Lett.}, vol. 25, no. 7, pp. 2395--2399, Jul. 2021.
	
	
	\bibitem{i2}
	K. Shen and W. Yu, ``Fractional programming for communication 	systems—part I: Power control and beamforming,'' \textit{IEEE Trans. Signal Process.}, vol. 66, no. 10, pp. 2616--2630, May 2018.
	
	\bibitem{i3}
	J. B. Moore and Danchi Jiang, ``A rank preserving flow algorithm for quadratic optimization problems subject to quadratic 	equality constraints,'' in \textit{Proc. IEEE ICASSP}, 1997, pp. 67--70.
	
	\bibitem{i4}
	Y. Cai, Z. Wei, R. Li, D. W. K. Ng, and J. Yuan, ``Joint trajectory and resource allocation design for energy-efficient secure UAV communication systems,'' \textit{IEEE Trans. Commun}., vol. 68, no. 7, pp. 4536--4553, Jul. 2020.
	
	\bibitem{i4-1}
	D. Li, ``Bound analysis of number configuration for reflecting elements in IRS-assisted D2D communications,'' \textit{IEEE Wireless Commun. Lett.}, vol. 11, no. 10, pp. 2220--2224, Oct. 2022.
	
	\bibitem{i5}
	E. Björnson, Ö. Özdogan, and E. G. Larsson, ``Intelligent reflecting surface versus decode-and-forward: How large surfaces are needed to beat relaying?,'' \textit{IEEE Wireless Commun. Lett.}, vol. 9, no. 2, pp. 244--248, Feb. 2020.
\end{thebibliography}
\end{document}